\documentclass[prb,showpacs,twocolumn,floatfix, groupaddress,superscriptaddress,footinbib]{revtex4-1}
\usepackage{times,amsmath,amsfonts,amssymb,latexsym}
\usepackage{graphicx,epsf}
\usepackage{color}
\usepackage{comment}
\usepackage{dsfont}
\usepackage{cancel}
\usepackage[toc,page]{appendix}

\newcommand{\bra}[1]{\langle #1 |}
\newcommand{\ket}[1]{| #1 \rangle}
\newcommand{\be}{\begin{equation}}
\newcommand{\ee}{\end{equation}}
\newcommand{\ba}{\begin{eqnarray}}
\newcommand{\ea}{\end{eqnarray}}
\DeclareMathOperator{\Tr}{Tr}
\newcommand{\ignore}[1]{}
\newcommand{\tens}[1]{%
  \mathbin{\mathop{\otimes}\limits_{#1}}%
}

\newcommand{\Id}{\mathds{1}}

\usepackage{hyperref}

\usepackage{graphicx}
\usepackage{amsmath}
\begin{document}

\title{Optimal parent Hamiltonians for time-dependent states}

\author{Davide Rattacaso }
\affiliation{Dipartimento di Fisica, Universit\`a di Napoli ``Federico II'', Monte S. Angelo, I-80126 Napoli, Italy}
\author{Gianluca Passarelli}
\affiliation{Dipartimento di Fisica, Universit\`a di Napoli ``Federico II'', Monte S. Angelo, I-80126 Napoli, Italy}
\affiliation{CNR-SPIN, c/o Complesso di Monte S. Angelo, via Cinthia - 80126 - Napoli, Italy}
\author{Antonio Mezzacapo}
\affiliation{IBM Quantum, IBM T.J. Watson Research Center, Yorktown Heights, NY 10598, USA}
\author{Procolo Lucignano}
\affiliation{Dipartimento di Fisica, Universit\`a di Napoli ``Federico II'', Monte S. Angelo, I-80126 Napoli, Italy}
\author{Rosario Fazio}
\affiliation{The Abdus Salam International Centre for Theoretical Physics , Strada Costiera 11, I-34151 Trieste, Italy}
\affiliation{Dipartimento di Fisica, Universit\`a di Napoli ``Federico II'', Monte S. Angelo, I-80126 Napoli, Italy}

\begin{abstract}
Given a generic time-dependent many-body quantum state, we determine the associated parent Hamiltonian. This procedure may require, in general, interactions of any sort. Enforcing the requirement of a fixed set of engineerable Hamiltonians, we find the optimal Hamiltonian once a set of realistic elementary interactions is defined. We provide three examples of this approach. We first apply the optimization protocol to the ground states of the one-dimensional Ising model and a ferromagnetic $p$-spin model but with time-dependent coefficients. We also consider a time-dependent state that interpolates between 
a product state and the ground state of a $p$-spin model. We determine the time-dependent optimal parent Hamiltonian for these states and analyze the capability of this Hamiltonian of generating the state evolution. Finally, we discuss the connections of our approach to shortcuts to adiabaticity.

\end{abstract}

\maketitle

\section{Introduction}
\label{introduction}

In the last decades, impressive progresses have been achieved in the realization and manipulation of a vast variety of engineered quantum 
systems~\cite{zoller2005}.  Solid-state quantum circuits, trapped ions, photonic or atomic/molecular systems are among the most successful 
examples of this kind~\cite{buluta2011}. In these implementations, control has reached levels that allow for an accurate driving of quantum states 
by means of a properly tailored (time-dependent) Hamiltonian. The essence of  most of quantum information protocols consists in a well 
defined time modulation of a set of couplings characterizing the Hamiltonian so to reach ultimately a final state that encodes the solution to 
the quantum task~\cite{NielsenChuang}. The modulation may break down into a set of elementary gates, evolve gently as in the adiabatic 
computation or possibly evolve through some global controls as in quantum simulators. In these situations, the Hamiltonian is known while the quantum states are not (the final one embeds the desired solution).

Differently from these cases, there are  some important instances in quantum information processing where one is interested in the inverse problem. Given a known 
quantum  (many-body) state, one would like to determine the {\em parent Hamiltonian} that generates it. What is the relevance of the inverse problem? 
In a nutshell, there are several many-body states that one wants to realize because of their properties, as for example Laughlin states\cite{Laughlin}, and would like to
understand how to generate them. The solution to this question is not unique. In general, given the state of the system, several solutions to 
the inverse problem exist in the space of Hermitian operators, but only a limited number of these Hamiltonians can be realized in a laboratory either 
due to fundamental constraints, like locality, or because of technological limitations.  For this reason, one usually considers solutions in a constrained space  of Hamiltonians. 

In the  time-independent case, the inverse problem consists in reconstructing  a stationary local Hamiltonian starting from a stationary 
state, for example an eigenstate. Following the pioneering work by Chertkov and Clark~\cite{Clark}, several works have been devoted to this 
question~\cite{Thomale,Zeng,Laflorencie,Ranard,lindner_stationary}. A key role in the determination of the parent Hamiltonian is played by the so-called  \emph{Quantum Covariance Matrix} (QCM) of the state. This matrix encodes the covariances of the local operators in the stationary state and its kernel contains all the possible local parent Hamiltonians. This approach has been extended in several ways, as  the search for a stationary parent Hamiltonian 
associated to a state after a quantum quench\cite{Hsieh}, to open quantum systems governed by a Lindblad dynamics~\cite{Lougovski,Arad} and the  determination of  a parent Hamiltonian of a Matrix Product State\cite{Monthus}. The non-uniqueness of the parent Hamiltonian is reflected in the  degeneracy of the kernel of the QCM. The characteristics of this degeneracy have been investigated~\cite{Ranard} showing under which conditions, 
as a consequence of the analyticity of the matrix, the degeneracy is removed and the reconstruction of the parent Hamiltonian from a unique eigenstate is possible. 
The possibility to reconstruct the parent Hamiltonian from measurements in a bounded region of space, and therefore from partial knowledge of the correlation 
matrix, has been discussed in Refs.~\onlinecite{Ranard,lindner_stationary}. A different approach to the inverse problem has been proposed~\cite{Dalmonte1} ,
based on the Bisognano-Wichmann theorem in quantum field theory, which relates the parent Hamiltonian to the reduced density matrix of the state. This method 
has been exploited to systematically reconstruct an approximate local parent Hamiltonian for Jastrow-Gutzwiller wave functions~\cite{Dalmonte2}.

In this work, we would like to consider the inverse problem for a time-dependent state. Given a quantum state $| \psi(t) \rangle$, we would like to determine the  
Hamiltonian $\hat H(t)$ that has $| \psi \rangle$ as a solution to the time-dependent Schr\"odinger evolution. The approach that we follow can be thought as a  time-dependent extension of the method proposed in Ref.~\onlinecite{Clark}. In particular, we will see that the QCM has a central role also in 
the time-dependent case. The reasons for considering this problem are multiple and, we believe, relevant in quantum information processing.  The most natural 
application is quantum state preparation. Given a possibly complex target state, it is possible to reach it by a time-path of quantum states once the Hamiltonian 
that generates this sequence of states is found.  An interesting possibility of this strategy, beyond the scope of the present work, is to design an inverse quantum 
adiabatic protocol for which the parent Hamiltonian can be designed by a quantum computer.  Also in this case, one needs to develop an approach to finding the 
most suitable Hamiltonian to perform this task taking into account the constraint on the available couplings present in the Hamiltonian (for example the range or the many-body character of the interactions).  In these terms, the question is closely similar to a quantum optimal control problem~\cite{Gross,opt_c_1,opt_c_2,opt_c_3,opt_c_4,opt_c_5,opt_c_6} , with two fundamental differences: here the target state is defined at each time and not only at the final time, and we consider an arbitrary space of engineerable interactions. Moreover, given an 
effectively realized state evolution, one could use this evolution to reconstruct the interactions affecting the system, taking advantage of the knowledge of the 
physical constraints affecting the Hamiltonian. This is time-dependent Hamiltonian learning~\cite{h_learning_1,h_learning_2,h_learning_3,h_learning_4}. Interesting connections of what we discuss here  can also be found with quantum verification. For a time-independent Hamiltonian, its relationship with inverse problems has been outlined~\cite{Zoller_verification,verification_1,verification_2,verification_3,verification_4,verification_5,verification_6}. Time-dependent Hamiltonian learning, 
quantum verification and optimal control are therefore different faces with a common ground, that is the search for a parent Hamiltonian for the 
state in a space of constrained Hermitian operators.

Since an exact parent Hamiltonian for a generic state usually does not satisfy the unavoidable constraints dictated by fundamental principles or by experimental 
requests, we look for an optimal parent Hamiltonian as the minimum of a suitable cost functional, in the sense 
that the fidelity between the target state and the state generated via the driving with a Hamiltonian goes to one when the associated cost goes to zero. 
This ensures that a low-cost solution can be useful both for quantum driving and for Hamiltonian learning. A fundamental point of our analysis is the direct geometrical 
meaning of the proposed cost functional, that reflects the geometry induced on the space of states by a set of allowed interactions. Interpreting the associated cost 
function as a geometrical object, we obtain the analytical expression of the optimal parent Hamiltonian at each time as a function of the state and its time derivative. 
The fundamental role of the QCM in this expression also demonstrates the link between the several proposals for stationary parent Hamiltonian 
reconstruction and the optimization methods involved in optimal control. Moreover, as we show in the body of this work, this geometrical point of view sets further links
between quantum control and the  questions regarding the accessibility of the Hilbert space~\cite{illusion_hilbert} and the geometrization of quantum 
complexity~\cite{Nielsen1,Nielsen2,susskindcomplexity}, allowing to design a geometrical picture of the 
capability of reproducing a given evolution with a fixed set of allowed interactions.

The paper is organized as follows. Section~\ref{t-dependent} is devoted to setting the stage by defining the time-dependent inverse problem. In this section, we show 
how to exploit the Euclidean structure of the space of Hermitian operators to find all the exact parent Hamiltonians for a time-dependent state. Although the inverse 
problems admit many solutions, engineering all the required interactions can be an impossible task. One is 
left with the problem of finding the parent Hamiltonian  in a restricted parameter space. In Section~\ref{optimal-ham}, we include the constraints on the space of 
achievable interactions by defining a cost functional that bounds the fidelity between the target time-dependent state and the evolution 
generated by a trial Hamiltonian. The minimization of this functional provides the optimal parent Hamiltonian. We show that the cost functional is locally related to the infinitesimal Fubini-Study length~\cite{Wootters} of the target path of states 
by an \emph{accessibility angle}. The latter defines a geometrical interpretation of our ability to generate a target evolution exploiting only some achievable interactions. In Section~\ref{examples}, we exploit our method to define an optimal parent Hamiltonian for several time-dependent states.
As will be clear in the following, the search for a  parent Hamiltonian related to a time-dependent quantum state has also tight links to shortcuts to 
adiabaticity~\cite{shortcuts1,shortcuts2,shortcuts3,shortcuts4,shortcuts5} performed via counterdiabatic driving\cite{Berry,counter_2,counter_3,Polkovnikov}. These links, as well as the effectiveness of the optimal parent Hamiltonian as a method to generate a shortcut to  adiabaticity, are discussed in Section~\ref{optimal-driving-ground}. 
The last part of the paper, Section~\ref{conclusions}, is devoted to conclusions and outlook.

\section{Time-independent inverse problem}
\label{t-dependent}

The problem of determining the exact parent Hamiltonian associated to a time-dependent quantum state can be easily posed. Given a pure time-dependent state 
$\hat{\rho}(t) = \ket{\psi(t)}\bra{\psi(t)}$ of a (possibly many-body) quantum system in the time interval $[0,T]$, the evolution of this state is related to the parent Hamiltonian $\hat H_\text{p}(t)$ via the Schr\"{o}dinger equation
\be\label{schr}
	\partial_t \hat{\rho}(t)=-i[\hat{H}_\text{p}(t),\hat{\rho}(t)].
\ee
The goal is to find $\hat{H}_\text{p}(t)$, given the state $\hat{\rho}(t)$. The exact parent Hamiltonian is not unique. Hence, to find all the solution to the inverse problem, we 
need to invert the Schr\"{o}dinger equation. This task becomes easier if we represent both the state and the unknown Hamiltonian as vectors in the space of Hermitian 
operators. This space, endowed with the Hilbert-Schmidt scalar product $\Tr(\hat{A}\hat{B})$, is  a Euclidean space, so we can choose an orthogonal basis 
$\{\hat{O}_\alpha\}$. As an example, an orthogonal basis for the Hermitian operators acting on an $N$-spins Hilbert space is given by all the tensor product operators $\hat O_\alpha =\hat  \sigma_{1,\mu_\alpha}\otimes \hat\sigma_{2,\nu_\alpha}\otimes...\otimes\hat\sigma_{N,\tau_\alpha}$, where $\hat\sigma_{i,\mu}$ is the $\mu^\text{th}$ Pauli matrix describing the $i^\text{th}$ spin ($\hat\sigma_{i,0}$ is the normalized identity $\hat{\mathds{1}}$ acting on the $i^\text{th}$  spin ).

Using this basis,  the quantum state 
$\hat\rho$ can be written as
\be
\hat{\rho}(t)= \sum_{\alpha} o_{\alpha}(t) \hat{O}_{\alpha},
\ee
where $o_{\alpha}(t)=\Tr (\hat{\rho}(t)\hat{O}_{\alpha})$, and a generic  Hamiltonian can be written as
\be
\hat{H}(t)=\sum_{\alpha} h_{\alpha}(t) \hat{O}_{\alpha},
\ee
where $h_\alpha(t)=\Tr (\hat{H}(t)\hat{O}_{\alpha})$.

At this point, we can project Eq.~(\ref{schr}) on each element of the basis $\{\hat{O}_\alpha\}$ through the Hilbert-Schmidt scalar product, obtaining a vector representation of the Schr\"{o}dinger equation in the form
\be\label{inv_sch}
\partial_to_{\alpha}(t)=- \sum_{\beta} h_{\beta}K_{\alpha,\beta}(t),
\ee
where $K_{\alpha,\beta}= i\sum_{l} o_{\gamma}(t)\Tr([\hat{O}_{\beta},\hat{O}_{\gamma}]\hat{O}_{\alpha})$ is named \emph{commutator matrix}.
At each time $t$, each exact parent Hamiltonian corresponds to a solution to this system of inhomogeneous equations for the coefficients $h_{\alpha}(t)$. The kernel $\text{Ker}(K)$ of the commutator matrix contains all the symmetries of the state $\hat\rho(t)$, that is, all the Hamiltonians that do not generate any evolution acting on this state. As a consequence of Eq.~(\ref{inv_sch}), the exact parent Hamiltonians, which form a vector space, differ from each other in terms of the elements of $\text{Ker}(K)$. In the stationary case, the kernel of $K_{\alpha,\beta}$ contains the solution to the stationary inverse problem and corresponds to the kernel of the QCM, as 
discussed in Ref.~\onlinecite{lindner_stationary}. In the next section, we will show how this matrix also determines the behavior of the optimal solutions 
to the time-dependent inverse problem. Here, instead, we just outline that the commutator matrix and the QCM respectively express the dynamical and the static 
aspect of an exact parent Hamiltonian for a stationary state: an operator lies in the kernel of $K_{\alpha,\beta}$ if it does not generate any dynamics, and lies in the kernel of the correlation 
matrix if it has zero variance.

As a simple illustration of what we discussed so far, it is useful to consider the parent Hamiltonian reconstruction for a time-dependent  spin-1/2 system. We consider, for 
example, the state
\be
\hat{\rho}(t)=(\sin(\omega t) \hat{\sigma}_x+\cos(\omega t) \hat{\sigma}_y + \hat{\mathds{1}})/2\nonumber
\ee
expressed in the orthogonal basis defined by the Pauli operators $\{\hat{\mathds{1}},\hat\sigma_x,\hat\sigma_y,\hat\sigma_z\}$. If we write the exact parent Hamiltonian as $\hat H_\text{p}=h_0(t)\hat{\mathds{1}}+h_x(t)\hat\sigma_x+h_y(t)\hat\sigma_y+h_z(t)\hat\sigma_z$, Eq.~(\ref{inv_sch}) reads 
\be
\begin{pmatrix}
	0\\
	\omega\cos(\omega t)\\
	\!-\omega\sin(\omega t)\!\\
	0\\
\end{pmatrix}
\!=\!\begin{pmatrix}
	0 & 0 & 0 & 0\\
	0 & 0 & 0 & 2 \cos(\omega t)\\
	0 & 0 & 0 & \!-2\sin(\omega t)\!\\
	0 & -\cos(\omega t) & \sin(\omega t) & 0\\
\end{pmatrix}
\!\!\begin{pmatrix}
	h_0\\
	h_x\\
	h_y\\
	h_z\\ 
\end{pmatrix}\nonumber
\ee
and the parent Hamiltonian for this evolution is
\be\label{oneSpinParent}
\hat{H}_\text{p}(t)=\frac{\omega}{2} \hat{\sigma}_z+f(t)[\sin(\omega t)\hat{\sigma}_x+\cos(\omega t)\hat{\sigma}_y]+g(t)\hat{\mathds{1}}
\ee
with $f(t)$ and $g(t)$ arbitrary functions of time. In the previous equation, the first term is the only one determining the evolution of the system. The second term is 
the arbitrary element of the kernel of the commutator matrix, which corresponds to the exact parent Hamiltonian for the steady state $\hat{\rho}(t)$ at each time and does 
not generate any evolution. The above example is also useful to understand the relationship between the inverse problem and shortcuts to adiabaticity. From this
perspective, the second term is the adiabatic Hamiltonian for $\hat\rho(t)$ while the first one is the counterdiabatic potential.

In general, for a many-body system the number of  equations in Eq.~(\ref{inv_sch}) grows exponentially with 
the size of the system. However, we will focus on realistic Hamiltonians containing only a fraction of the operators $\hat{O}_{\alpha}$ implying few-body interactions,  with well behaved couplings. That could be well implemented in experimental systems. 

The optimization method necessary to construct effective Hamiltonians with few physically relevant interactions is illustrated in the next Section.

\section{Optimal parent Hamiltonian}
\label{optimal-ham}

In the previous section, we have shown that all the exact solution to the inverse problem  are solution to Eq.~(\ref{inv_sch}). 
Physically relevant systems  can be described by  linear superpositions of a limited number ``elementary" interactions $\{\hat L_a\}$. Here, as in the rest of this paper, we use the letter $L$ for operators in set of the \emph{realistic} elementary interactions, while the letter $O$ is associated to operators that do not have to satisfy any constraint. The form of these elementary interactions depends on  physical and technological constraints.  Usually this subset includes a polynomial (in the size of the system) number of operators. 
For instance, for  spin-1/2  lattices, a suitable basis of local interactions is the set 
$\{\hat{\sigma}_{i,\mu}\}\cup\{\hat{\sigma}_{i,\mu}\otimes\hat{\sigma}_{j,\nu}\}\cup\{\hat{\sigma}_{i,\mu}\otimes\hat{\sigma}_{j,\nu}\otimes\hat{\sigma}_{l,\tau}\}$ 
of one-, two-, and three-body operators between the lattice sites $i,j,l$. The Hamiltonian is the span of these operators and  we define it as $k$-local, if the maximum distance  between interacting sites is $k=\max\{|i-j|,|j-l|,|i-l|\}$. In the following, we will often suppose that the constraint that determines this basis is locality, but our considerations can be easily extended to arbitrary sets 
of allowed interactions.

\begin{figure}[htp]
\centering
\includegraphics[scale=1]{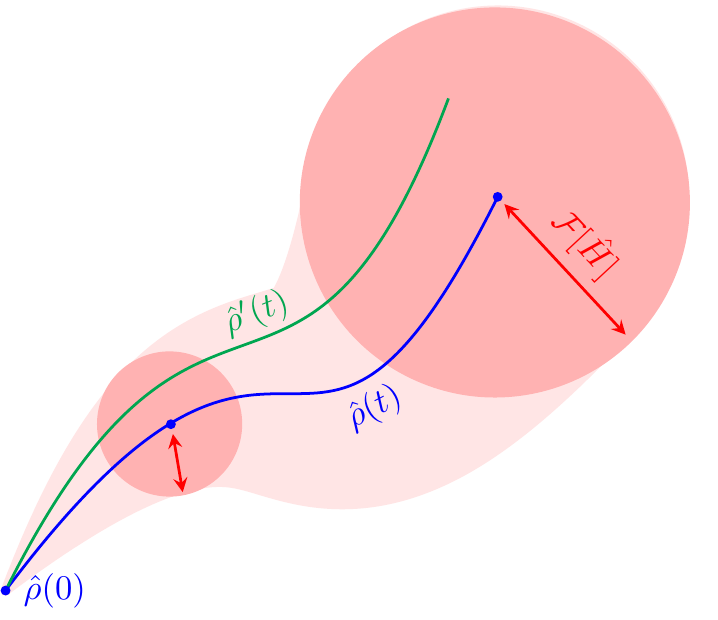}
\caption{At each time $t$, the state $\hat\rho(t)$ generated by the Hamiltonian $\hat H(t)$ lies inside a ball of radius $\mathcal{F}[\hat H,t]$ centered on the target state $\hat \rho(t)$.}
\label{distance_bound}
\end{figure}

We express  a generic Hamiltonian with locality constraint as  $\hat{H}(t)=\sum_{i} h_{a}(t) \hat{L}_{a}$. 
The set of the possible Hamiltonians, endowed with the Hilbert-Schmidt product defined in the previous section, forms a Euclidean subspace of the space of Hermitian operators. Therefore we can look for a realistic exact parent Hamiltonian by exploiting again the method defined in the previous section, that is by looking for solution to Eq.~(\ref{inv_sch}), where the commutator matrix has the form  $K_{a,\alpha} = i\sum_{\beta} o_{\beta}(t)\Tr([\hat{L}_{a},\hat{O}_\beta]\hat{O}_{\alpha})$. Since the cardinality of the basis of local operators is polynomial in the  size of the system,  while the dimension of the space of infinitesimal evolutions for a quantum state is exponential in its size, for an arbitrary time-dependent state we cannot expect that a realistic parent Hamiltonian exists.  Hence, we will seek for  optimal parent Hamiltonians as  approximate solutions to the inverse problem by minimizing the cost functional
\be
\label{cost_functional}
	\mathcal{F}[\hat H]=  {\int_0^Tdt f(\hat H,t)}= \int_0^Tdt\lVert\partial_t \hat{\rho}(t)+i[\hat{H}(t),\hat{\rho}(t)]\rVert \;,
\ee
where the argument of the last integral is the Frobenius norm $\lVert\cdot\rVert=\sqrt{\Tr(A^2)}$ of the difference between the RHS and the LHS of the Schr\"{o}dinger equation.
Of course, this is a measure of the error done in approximating the exact infinitesimal evolution $\partial_t \hat{\rho}(t)$ of the state with the evolution $-i[\hat{H}(t),\hat{\rho}(t)]$ 
generated by the candidate parent Hamiltonian $\hat H(t)$. Moreover, as shown in Appendix~\ref{appendix-bound}, $\mathcal{F}[\hat H]$ gives an  upper bound for the Hilbert-Schmidt distance 
between the target state $\hat \rho(T)$ and the state
\be\label{opt_ev}
\hat\rho_\text{opt}(T)=\left(\mathcal{T}e^{-i\int_0^T \hat H_\text{opt}(t)dt}\right)\hat\rho(0)\left(\mathcal{T}e^{-i\int_0^T \hat H_\text{opt}(t)dt}\right)^\dag
\ee
 evolved with the optimal parent Hamiltonian. As illustrated in Figure~\ref{distance_bound}, this means that the state $\hat \rho_\text{opt}(t)$, during its evolution, has to stay in a ball of radius $\mathcal{F}[ \hat H,t]=\int_0^tf(\hat H_\text{opt},t')dt'$ with center in $\hat \rho(t)$. In terms of the squared fidelity $F(\hat{\rho}(T),\hat{\rho}_\text{opt}(T))=\Tr(\hat{\rho}(T)\hat\rho_\text{opt}(T))$, which measures how much the state $\hat\rho_\text{opt}$ is experimentally indistinguishable from the state $\hat\rho$, inequality in Eq.~(\ref{opt_ev}) becomes
 \be
\label{fidelity_bound}
1\geq F(\hat{\rho}(T),\hat\rho_\text{opt}(T))\geq 1-\frac{1}{2}\mathcal{F}[\hat{H},T]^2.
\ee

Eq.~\eqref{cost_functional} assures that the cost functional $\mathcal{F}[\hat H]$ is minimized  when $\hat{H}(t)$ locally generates an evolution as close as possible to $\partial_t\hat{\rho}(t)$, while inequality in Eq.~\eqref{fidelity_bound} assures that a sufficiently low-cost minimum of $\mathcal{F}$ generates a high-fidelity driving of the initial state. However, that does not imply that the minimum of $\mathcal{F}$  is the best solution to the inverse problem in terms of fidelity: some Hamiltonians may exist, with larger cost,  generating the driving with higher fidelity.

Once defined the optimal parent Hamiltonian as the minimum of the functional $\mathcal{F}[\hat H]$ in Eq.~\eqref{cost_functional}, we can recast our problem from a  global to a local form by 
expressing the derivative of the Hamiltonian as a function of the state evolution. The integrand of Eq.~(\ref{cost_functional}) does not depend on the derivative of 
the Hamiltonian, therefore a continuous minimum for $\mathcal{F}[\hat H]$ can be obtained as the time-dependent minimum of the integrand $f(\hat H,t)$. In other words, the 
time-dependent state determines a time-dependent potential and the Hamiltonian evolves in time remaining in the minimum of this potential, as in an adiabatic system.

\begin{figure}[htp]
\centering
\includegraphics[scale=.700]{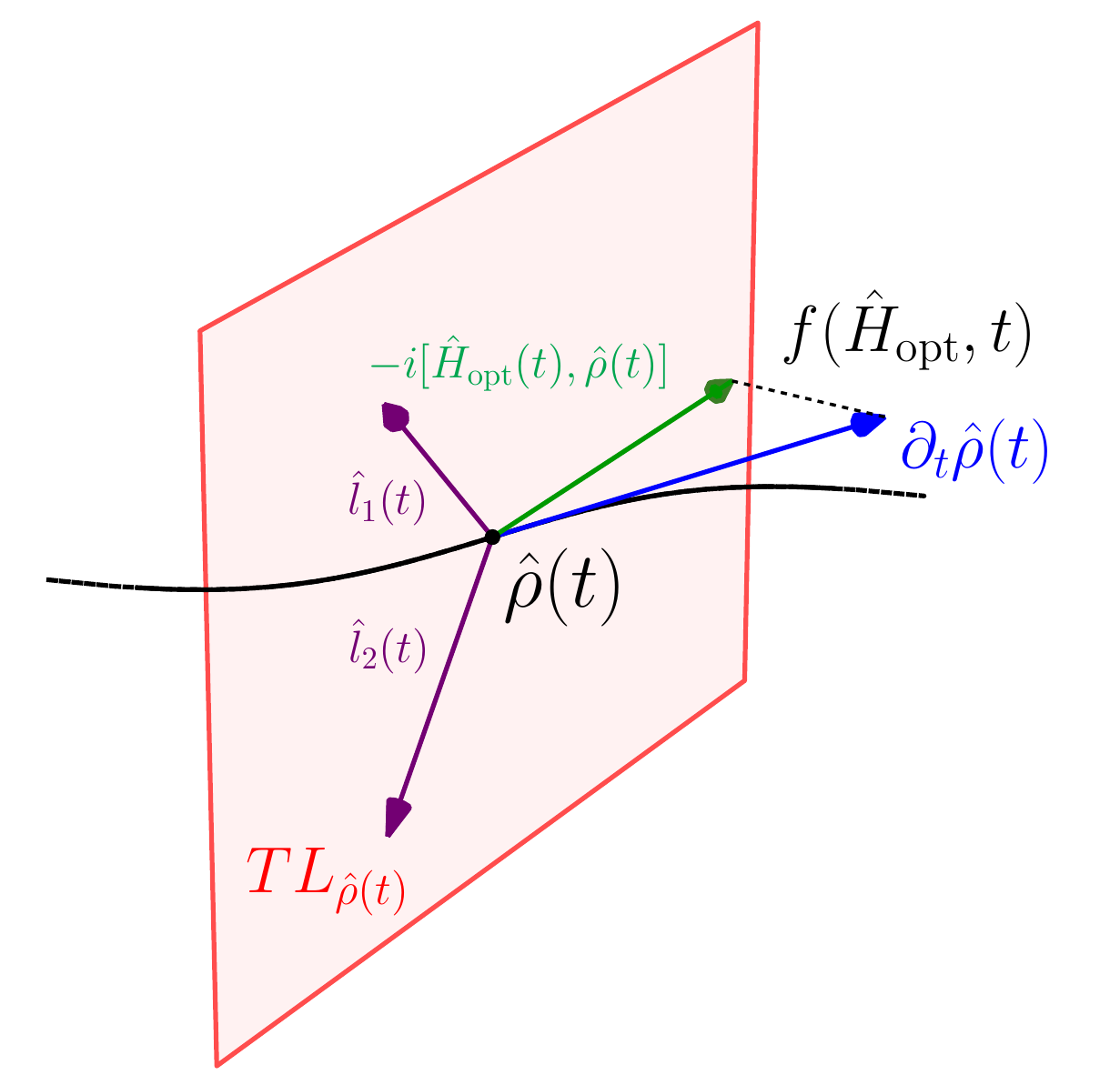}
\caption{The commutator $-i[\hat{H}_\text{opt}(t),\hat\rho(t)]$ has to be the projection of $\partial_t\hat\rho(t)$ on the space spanned by the vectors $\hat l_a(t)$.}
\label{TLpsi}
\end{figure}

In Figure~\ref{TLpsi},  we show a geometrical  interpretation of $f(\hat H,t)$. The time-derivative $\partial_t \hat\rho$ of the state $\hat\rho(t)$ is a vector in the 
tangent space $T_{\hat \rho(t)}$ for the state $\hat\rho(t)$, that is, the space of all the infinitesimal unitary evolutions and is spanned by the vectors $\{-i[\hat O_\alpha,\hat\rho(t)]\}$. On the other hand, the term $i[\hat H,\hat\rho(t)]$ is constrained in the subspace $TL_{\hat\rho(t)}$ of $T_{\hat\rho(t)}$ 
spanned by the vectors  $\hat l_a(t)= -i [\hat L_a,\hat{\rho}] $, that is, the space of evolutions that can be generated by local operators. The cost $f(\hat H,t)$ is the Euclidean (Hilbert-Schmidt) distance between the vectors $\partial_t\hat\rho(t)$ and $-i[\hat H_{\text{opt}}(t),\hat \rho(t)]$, hence its 
minimum  is given by the Hamiltonian $\hat{H}_\text{opt}(t)$ such that $-i[\hat{H}_\text{opt}(t),\hat{\rho}(t)]$ is the projection of $\partial_t \hat\rho(t)$ on $TL_{\hat\rho(t)}$.
When $i[\hat{H}_\text{opt}(t),\hat{\rho}(t)]$ is the projection of $\partial_t \hat\rho(t)$, the projections of these two vectors on the $\hat l_a(t)$ coincide. In formulas, this means that
\be\label{main_0}
\Tr(\hat l_a(t) \partial_t \hat{\rho}(t))=-i\Tr(\hat l_a(t) [\hat{H}_\text{opt}(t),\hat{\rho}(t)])
\ee
for each $\hat l_a$. If we replace the expression $\hat H_\text{opt}(t)=\sum_i h_{\text{opt},a}(t)\hat L_a$ of the optimal parent Hamiltonian in terms of the couplings in Eq.~(\ref{main_0}), it becomes
\be\label{main_eq}
\Tr(\hat l_b(t)\partial_t \hat{\rho}(t))=\sum_a\Tr(\hat l_a(t) \hat l_b(t))h_{\text{opt},b}(t)
\ee
where $\Tr(\hat l_a(t) \hat l_b(t))$ corresponds to the QCM
\ba
V_{ab}(t)&=&\bra{\psi(t)}\{\hat{L}_a ,\hat{L}_b\}\ket{\psi(t)}\nonumber\\
&-&2\bra{\psi(t)}\hat{L}_a\ket{\psi(t)}\bra{\psi(t)}\hat{L}_b\ket{\psi(t)}\nonumber
\ea
introduced by Chertkov and Clark\cite{Clark} in the context of the inverse problem associated to a time-independent state.
Since the vectors $\hat l_a(t)$ are generally not linearly independent, the Gram matrix $V_{ab}(t)$ is not invertible. In particular, as shown in Ref.~\onlinecite{Clark}, the kernel of this matrix contains all the symmetries of the quantum state at a given time. As a consequence, Eq.~(\ref{main_eq}) represents a linear inhomogeneous system 
with several solutions, that are the minima of the cost function $f(\hat H,t)$ at time $t$. Even if all these solutions have the same cost $f$, they will generate a different time evolution with a different fidelity between the evolved state and the target state. In the rest of this paper, we will focus our attention on the simplest solution to Eq.~(\ref{main_eq}), that is, the one that does not include the instantaneous 
symmetries contained in the kernel of the QCM. 

This particular choice has an important consequence on the fidelity  between the target time-dependent state $\hat\rho(t)$ and the state $\hat\rho_\text{opt}(t)$ generated by the optimal parent Hamiltonian. Indeed, if we define a new time-dependent state $\hat\phi(\tau)=\hat\rho(t(\tau))$ evolving through the same path but at different velocity, we have that
\be
F(\hat\phi(\tau(t)),\hat\phi_\text{opt}(\tau(t)))=F(\hat\rho(t),\hat\rho_\text{opt}(t)),
\ee
where $\hat\phi_\text{opt}(\tau(t))$ is the evolution generated by the optimal parent Hamiltonian associated to $\hat\phi(\tau)$. In words, with this choice of the optimal parent Hamiltonian, the fidelity does not depend on the velocity of the target evolution. This is a consequence of the fact that the solution to Eq.~\eqref{main_eq} that does not include the instantaneous symmetries is proportional to the time derivative of the state: an operator $\hat A(\lambda)$ exists such that $\hat H_\text{opt}(t)=\dot\lambda(t) \hat A(\lambda)$, hence the generated unitary evolution $\hat U(t)=\mathcal{T}\exp\left(-i\int_0^t \dot\lambda(t') \hat A(\lambda)dt'\right)$ only depends on $\lambda(0)$ and $\lambda(t)$. Note that if we include the element of the kernel of the QCM in the optimal Hamiltonian, it is not anymore proportional to the derivative of the state and the previous statement is not valid.

We conclude this section with a comment on the physical meaning of the angle between the vectors $\partial_t\hat\rho(t)$ and $-i[\hat H_\text{opt}(t),\hat\rho(t)]$. In Figure~\ref{TLpsi} we can see that
the optimal cost is proportional to the Fubini-Study length of the evolution $\partial_t\hat\rho(t)$ through the sine of this angle. Therefore, this angle, that we call \emph{accessibility angle}, determines the capability of 
the allowed interactions to generate the target time evolution. The more difficult it is to access the target evolution $\hat\rho(t)$ using the available interactions $\{\hat L_a\}$, the 
closer the accessibility angle is to $\pi/2$ during the evolution. Since the difficulty to access quantum states with a finite amount of resources and the time necessary to 
perform this task are two sides of the same coin, we suggest that this angle could have a central role in the geometrization of quantum complexity proposed by Nielsen 
in the context of unitary operators and extended by Susskind to the space of quantum states~\cite{Nielsen1,Nielsen2,susskindcomplexity}. Finally, this relation between 
cost and length of the path helps us understand to what extent it is possible to exploit the geodesics of the Fubini-Study metric to select the most suitable path of 
states to prepare a given final quantum state~\cite{Brachistochrone,AdiabBrachistochrone}.

\section{Examples}
\label{examples}

In this section, we apply the general approach illustrated so far to some specific examples.  We  discuss three cases in detail. In the first two examples, the time-dependent state $\hat\rho(t)$ coincides with the instantaneous ground states of given many-body Hamiltonians: the one-dimensional Ising model  and the $p$-spin model in a  time-dependent transverse field. It is interesting to analyze this type of dynamics since it relates to shortcuts to adiabaticity.  In the  third example, we  consider a generic case in which two ground states are connected through an arbitrary sequence of states.

\subsection{Time-dependent ground state of the Ising model in transverse field}
\label{ising_example}

We start from a paradigmatic example, and consider the state $\hat\rho(t)=\ket{\psi(t)}\bra{\psi(t)}$ as the instantaneous ground state of the one-dimensional 
Ising model in a time-dependent transverse field, i.e.,  $\hat H_\text{I}(\lambda(t)) \ket{\psi(t)} = E_\text{GS} (t) \ket{\psi(t)}$ where $E_{\text{GS}} (t)$ is the energy of the instantaneous 
ground state and the Ising Hamiltonian in a time-dependent transverse field $\lambda(t)\in[0,3]$ is given by
\be
\hat H_\text{I}(\lambda(t))=-J\sum_{i=1}^L\hat\sigma_{i,x}\hat\sigma_{i+1,x}-\lambda(t)\sum_{i=1}^L\hat\sigma_{i,z} \; .
\ee
We consider periodic boundary conditions $\hat\sigma_{L+1,\mu}=\hat\sigma_{1,\mu}$ and an even number of sites $L$, moreover we set $J=1$. It is important to remark that the Hamiltonian 
$\hat H_\text{I}$ only determines the state $\hat \rho(t)$ for which we will solve the inverse problem, while it is not involved in the definition of the time-dependent parent Hamiltonian.
Then, starting from $\hat \rho(t)$ we obtain its optimal parent Hamiltonian  with local interactions  minimizing the cost  functional defined in Eq.~\eqref{cost_functional}.

First of all,  we  represent the Hamiltonian $\hat H_\text{I}(\lambda)$ in a diagonal form. Since the ground states lie in the even parity sector of the Hilbert space for the symmetry operator $\hat Q=\prod_{i=1}^L{\hat\sigma_{i,z}}$, we will represent the Hamiltonian only in this sector. Exploiting the equivalence of the Ising model in transverse field with a chain of non-interacting Anderson pseudo-spins~\cite{pseudo_spins} $\tilde\sigma_k^\mu$  associated to the momenta in $\mathcal{K}^\pm=\{\pm\frac{(2i-1)\pi}{L}\text{, with }i=1,...,L/2\}$ , we find the following representation of $\hat H_\text{I}(\lambda)$ (see Appendix~\ref{appendix-Ising}):
\ba
\hat H_\text{I}&=&\sum_{k\in\mathcal{K}^+}\hat H_k;\nonumber\\
\hat H_k&=&\epsilon_k\left(\cos(\theta_k)\tilde\sigma_k^z+\sin(\theta_k)\tilde\sigma_k^x\right);\nonumber
\ea
where $\epsilon_k(\lambda)=2\sqrt{(\lambda-\cos(k))^2+\sin^2(k)}$ and $\theta_k(\lambda)=-\arctan\left(\frac{\sin(k)}{\lambda-\cos(k)}\right)$. In this representation, the 
ground state is the tensor product of single pseudo-spin ground states of the Hamiltonians $\hat H_k$,
\be
\ket{\psi(t)}=\tens{k\in\mathcal{K}^+}\left(-\sin(\theta_k/2)\ket{\uparrow_k}+\cos(\theta_k/2)\ket{\downarrow_k}\right)\label{psi_ising},
\ee
that defines our path  $\hat \rho (t)$.
In  Eq.~(\ref{oneSpinParent}),  we have shown how to apply our method to reconstruct the  exact parent Hamiltonian for a single spin. We can generalize that result to the state of Eq.~\eqref{psi_ising} to find its exact parent Hamiltonian in the form of a the sum of the corresponding rotations of single pseudo-spins:
\be
\hat H_\text{p}(t)=\frac{1}{2}\dot\lambda\sum_{k\in\mathcal{K}^+}\left(\partial_\lambda\theta_k\right)\tilde\sigma_k^y.\label{BP}
\ee
Remarkably, in this case, the exact parent Hamiltonian is also the counterdiabatic potential for the Ising Hamiltonian $\hat H_{\text{I}}(\lambda)$\cite{Zurek}. This solution is not unique: there are  other exact solutions to the inverse problem for $\hat\rho(t)$. They  can be obtained from Eq.~\eqref{BP} by adding any term $\hat H_S(t)$ such that $[\hat H_S(t),\hat\rho(t)]=0$.

The Hamiltonian $\hat H_\text{p}(t)$ drives the initial state through the target evolution with fidelity one since it is an exact solution to the inverse problem. Expressing Eq.~\eqref{BP} in terms of real-space Pauli matrices $\hat\sigma_{i,\mu}$, we obtain
\be \label{isingParent}
\hat H_\text{p}(t)=\dot\lambda\sum_{j'>j}\omega_{j,j'}\mathcal{\hat S}_{jj'}\;,
\ee
where
\be
\omega_{j,j'}=\frac{1}{2L}\sum_{k\in\mathcal{K}^+}\partial_\lambda\theta_k\sum_{j'>j}\sin[k(j'-j)]\nonumber
\ee
and
\ba
\mathcal{\hat S}_{jj'}&=&\hat\sigma_{j,x}\hat\sigma_{j+1,z}\dots\hat\sigma_{j'-1,z}\hat\sigma_{j',y}\nonumber\\
&+&\hat\sigma_{j,y}\hat\sigma_{j+1,z}\dots\hat\sigma_{j'-1,z}\hat\sigma_{j',x}\nonumber
\ea
is a string of Pauli operators acting on the spins from $i$ to $j$, in which the dots represent a string of $\hat\sigma_z$ operators. Therefore, $\hat H_\text{p}$ is nonlocal and involves interactions on a number of spins that scales linearly with the size of the system.
 
So far, our approach allows for an exact solution in terms of arbitrary long strings of spin operators. 
Now we look for an approximate   solution to the inverse problem. In particular, we look for an optimal parent Hamiltonian for $\hat\rho(t)$ in the space spanned by 
single- and two-spin interactions in the set $B_L=\{\hat\sigma_{i,\mu}\}\cup\{\hat\sigma_{i,\mu}\hat\sigma_{i+1,\nu}\}$. To this end, we 
replace the derivative of the state $\partial_t\hat\rho$ in Eq.~(\ref{main_eq}) with $-i[\hat H_\text{p},\hat\rho]$. In this way, we obtain the filter function that relates 
the couplings of the optimal parent Hamiltonian to the couplings of the exact parent Hamiltonian. Here, it is important to remark that, since the different exact solution to the inverse problem generate the same infinitesimal evolution of the state, a different choice of the exact parent 
Hamiltonian has no effect on the commutator $[\hat H_\text{p},\hat\rho]$ in Eq.~(\ref{main_eq}) and therefore it leads to the same optimal parent Hamiltonian.

\begin{figure}[htp]
\centering
\includegraphics[scale=.3]{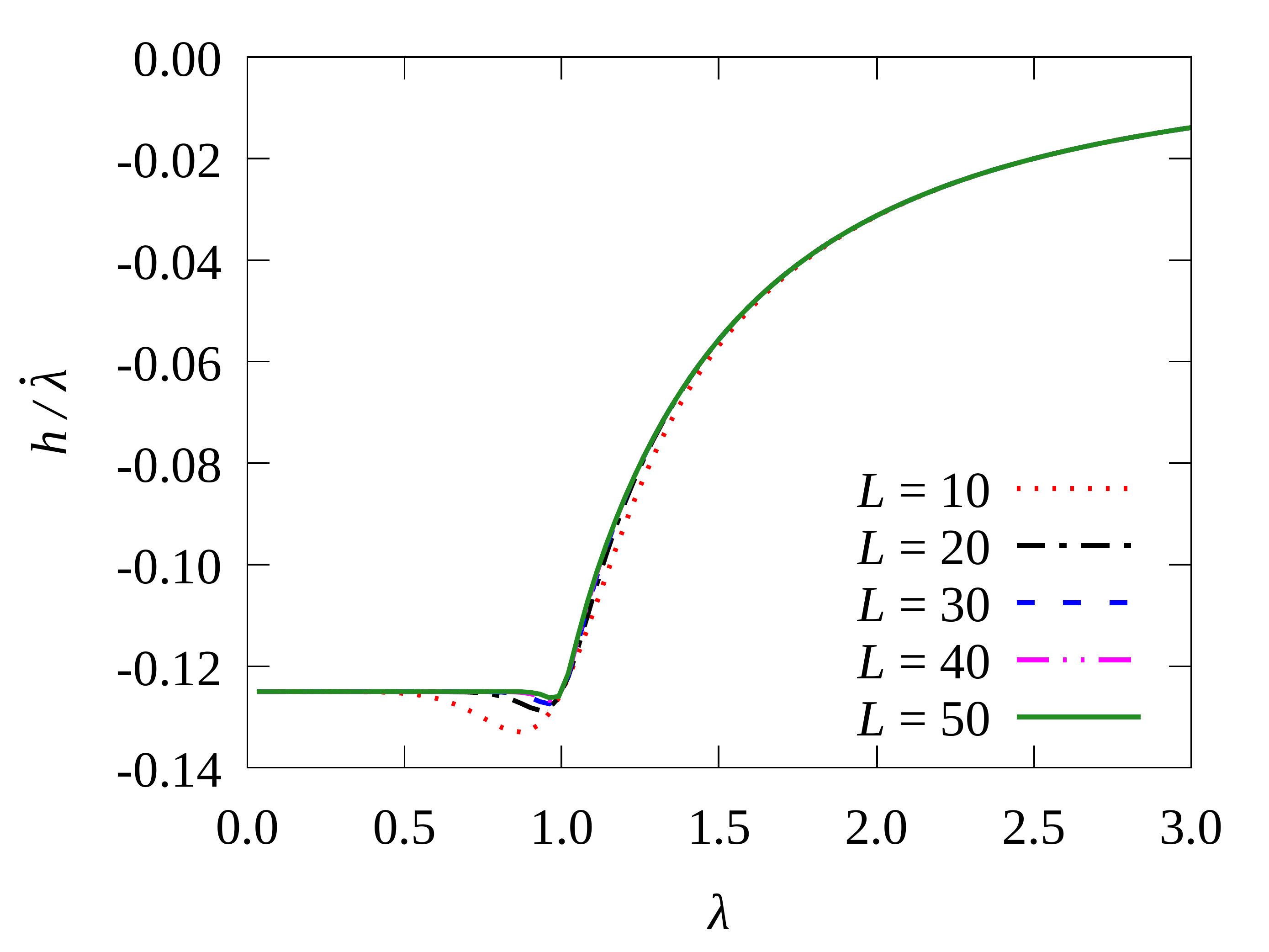}
\caption{Evolution of the (rescaled) optimal coupling $h/\dot\lambda$ for different system sizes. $\lambda\in[0,3]$ is an arbitrary smooth function of time.}
\label{ising_coupling}
\end{figure}

At this point, we use the symmetries of the problem to  simplify the resolution of Eq.~(\ref{main_eq}) (see Appendix~\ref{appendix-Ising}). In particular, we exploit the axis 
reflection symmetry of the Hamiltonian to show that the observables involving an odd number of $\hat\sigma_{i,x}$ or $\hat\sigma_{i,y}$ have zero expectation value. 
Moreover we exploit the symmetry of the ground state for a rotation $i\rightarrow i+1$ and for the reflection $i\rightarrow -i$ of the real-space axis. In this way, we prove 
that the optimal parent Hamiltonian has the form
\be\label{isingOptimal}
\hat H_\text{opt}=h(t)\sum_i(\hat\sigma_{i,x}\hat\sigma_{i+1,y}+\hat\sigma_{i,y}\hat\sigma_{i+1,x})
\ee
where
\be
h(t)=\dot\lambda\sum_{l'>l}\omega_{l,l'}\frac{\langle\mathcal{\hat S}_{ll'}\sum_i\mathcal{\hat S}_{ii+1}\rangle}{\langle \sum_i\mathcal{\hat S}_{ii+1}\sum_j\mathcal{\hat S}_{jj+1}\rangle}\label{h(t)1}
\ee
is the only nonzero coupling and $\langle \cdot\rangle$ is the expectation value of an operator on the target state $\hat\rho(t)$. The correlation functions  in Eq.~\eqref{h(t)1} have been previously computed~\cite{Pfeuty,Lieb} exploiting the exact diagonalization of the Hamiltonian $\hat H_\text{I}(\lambda)$ and  Wick's theorem. A compact  expression of the optimal coupling is obtained through the Anderson pseudo-spin representation:
\be
h(t)=-\dot\lambda\frac{\sum_{k,k'\in\mathcal{K}^+}\left\langle \left(\partial_\lambda\theta_k\right)\tilde\sigma_k^y\sin(k')\tilde\sigma_{k'}^y\right\rangle}{8\sum_{k,k'\in\mathcal{K}^+}\left\langle\sin(k')\tilde\sigma_{k'}^y\sin(k)\tilde\sigma_k^y\right\rangle}.\nonumber
\ee
Exploiting the orthogonality relation $\left\langle\tilde\sigma_{k'}^y\tilde\sigma_k^y\right\rangle=\delta_{k'k}$, this simplifies to
\be
h(t)=-\dot\lambda\frac{\sum_{k\in\mathcal{K}^+}\partial_\lambda\theta_k\sin(k)}{8\sum_{k\in\mathcal{K}^+}\sin(k)^2},\nonumber
\ee
which is the exact expression of the optimal coupling. This expression depends on the time schedule only 
through $\lambda(t)$ and its time derivative. The behavior of the renormalized coupling $h(t)/\dot\lambda$ for different values of $\lambda$ is shown in Figure~\ref{ising_coupling} 
for different system sizes. The magnitude of the couplings does not scale with the size of the system, allowing for a scalable implementation of the optimal Hamiltonian in synthetic quantum systems.

\begin{figure}[htp]
\centering
\includegraphics[scale=.3]{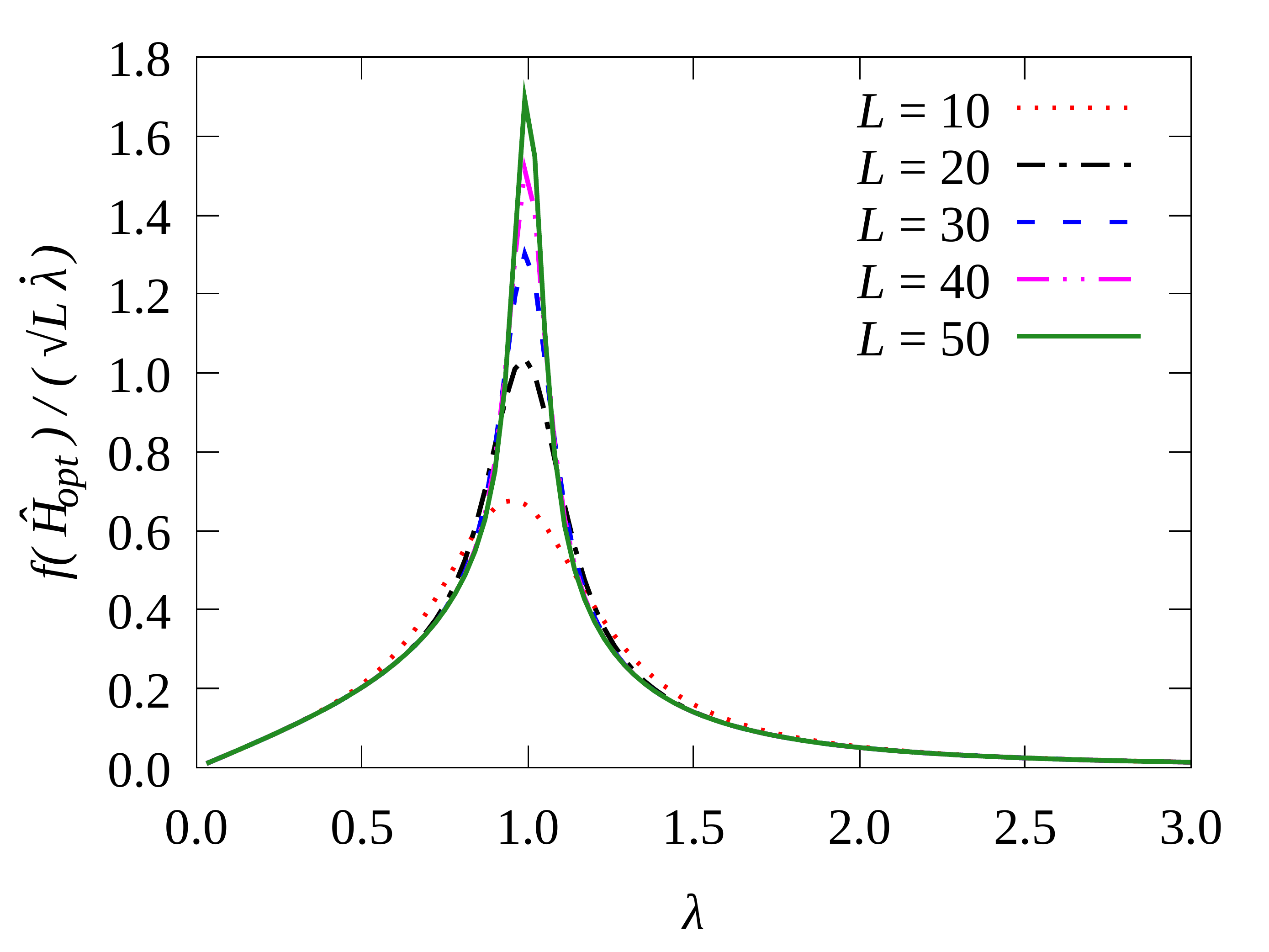}
\caption{Evolution of the (rescaled) optimal cost $f(\hat H_\text{opt})/(\sqrt{L}\dot\lambda)$ for different system sizes. $\lambda\in[0,3]$ is an arbitrary smooth function of time.}
\label{ising_cost}
\end{figure}

The rescaled local cost $f(\hat H_\text{opt},t)/(\sqrt{L}\dot\lambda)$ of the optimal solution is shown in Figure~\ref{ising_cost} for different system sizes. For noncritical states, the different curves converge for a sufficiently large system, signaling that $f(\hat H_\text{opt},t)$ scales with $\sqrt{L}$. For the critical states, instead, the cost scales faster than $\sqrt{L}$ and finding a local optimal parent Hamiltonian becomes very hard. This behavior for noncritical states is a direct consequence of the scaling of the Fubini-Study length of the target path of 
states~\cite{Zanardi1,Zanardi2}. This length indeed, as we can see from Figure~\ref{TLpsi}, upper bounds the cost of the optimal Hamiltonian.

\begin{figure}[htp]
\centering
\includegraphics[scale=.3]{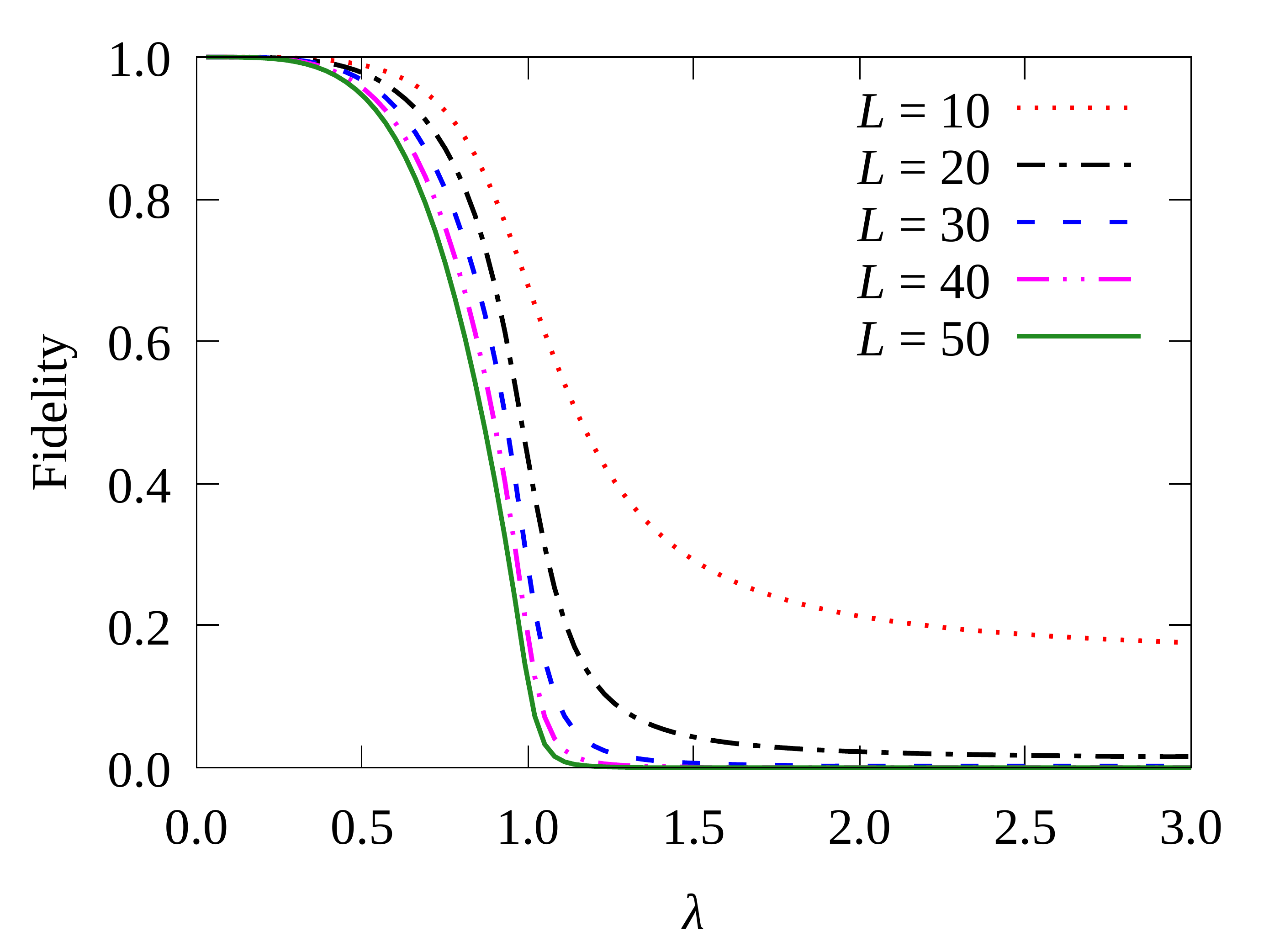}
\caption{Evolution of the fidelity between the target state and the state evolved through the optimal parent Hamiltonian for different system sizes. $\lambda\in[0,3]$ is an arbitrary smooth function of time.}
\label{ising_fid}
\end{figure}

The fidelity $F(t)=\Tr(\hat \rho(t)\hat \rho_p(t))$ between $\hat \rho(t)$ and the state $\hat \rho_p(t)$ driven by the optimal parent Hamiltonian assumes a very simple 
form in the pseudo-spin representation of the Hilbert space:
\be
F(t)=\prod_{k\in\mathcal{K}^+}|\bra{\psi(t_0)}e^{i\int_{t_0}^tdt'(\frac{1}{2}\dot\lambda\partial_\lambda\theta_k+4h(t)\sin(k))\tilde\sigma_k^y}\ket{\psi(t_0)}|^2.\nonumber
\ee
Defining the time-dependent angle $\alpha_k(t)=4\sin(k)\int_{t_0}^th(t')dt'+\frac{\theta_k(\lambda(t))-\theta_k(\lambda(t_0))}{2}$, we obtain
\be
F(t)=\prod_{k\in\mathcal{K}^+}\cos(\alpha_k(t))^2.\nonumber
\ee
As anticipated in the previous section, the fidelity depends on time through $\lambda$, but does not depend on the time schedule. In 
Figure~\ref{ising_fid}, the fidelity between the target state and the driven state is shown for different system sizes. Away from the critical coupling $\lambda_c=1$, the 
optimal parent Hamiltonian performs the driving of the state with high fidelity, while a large amount of fidelity is irreversibly lost in correspondence of the critical value 
of the control parameter. This behavior can be predicted from the shape of the cost in Figure~\ref{ising_cost} by virtue of the inequality of Eq.~(\ref{fidelity_bound}). The drastic drop of fidelity in correspondence of the critical state is a consequence of Lieb-Robinson bounds~\cite{LiebRobinson}. These bounds state that the local correlation length increases at a finite velocity under the evolution generated by a local Hamiltonian. Since this velocity is proportional to the magnitude of the interactions involved in the Hamiltonian, this means that if we want the correlation length to diverge in the thermodynamic limit, as it happens in the critical states, we need also diverging magnitude of the couplings. As we see in Figure~\ref{ising_coupling}, here the couplings do not scale with the size of the system, hence the critical state becomes inaccessible as predicted by the Lieb-Robinson bounds and we see a drastic drop of the fidelity between the target state and the state generated by the optimal Hamiltonian.

The problem of constructing the (exact or approximate) time-dependent Hamiltonian generating the state considered here through shortcuts to adiabaticity has been 
also discussed by del Campo, Rams and Zurek~\cite{Zurek} as an application of the counterdiabatic driving~\cite{Berry}. Their aim  was to understand the performance of a local counterdiabatic potential for the Ising model as a shortcut to adiabaticity. 
To this end, the authors have exploited the Ising diagonal form of the Hamiltonian to determine the counterdiabtic potential in Eq.~(\ref{isingParent}), and then have filtered out nonlocal interactions. This yields our Eq.~(\ref{isingOptimal}). This situation rises an important question: does our optimal parent Hamiltonian 
for time-dependent ground states generally correspond to the exact counterdiabtic potential in which nonlocal interactions have been removed? Our answer to this question is negative: the filter function that our optimization method implicitly defines is analyzed in Appendix~\ref{appendix_filter} where, from Eq.~(\ref{filter}), we can see that the filter depends on the effect of the interactions only on the target  state. On the other hand the procedure followed in Ref.~\onlinecite{Zurek} does not depend on the target states and therefore takes into account the effect of the interactions in the whole Hilbert space. This may become more evident in the following subsection.

\subsection{Time-dependent ground state of the $p$-spin model}

Another paradigmatic case is the inverse problem for the ground state of a time-dependent $p$-spin Hamiltonian in transverse field. This Hamiltonian, originally introduced in~\cite{derrida} and~\cite{mezard} to model spin glasses, consists in a spin system in which the elementary interactions connect each set of $p$ spins regardless of their distance. From the computational point of view, the nonlocal interactions involved in this model are required for the resolution of some computationally hard problems like the Grover's search\cite{grover}. Hence, the possibility of generating the dynamics induced by this kind of Hamiltonians exploiting realistic resources can give an interesting boost in the direction of quantum algorithm implementation. For this reason, several works have been dedicated in recent years to propose techniques to implement the ground states of the $p$-spin Hamiltonian, with a particular focus on techniques that involve shortcuts to adiabaticity~\cite{Passarelli1,Passarelli2,Passarelli3,Passarelli4,Nishimori,Santoro1,Santoro2,Santoro3}. The search for an optimal parent Hamiltonian can be considered a further contribution to the debate on this topic.

The Hamiltonian of the $p$-spin model, and consequently its ground state, can be easily represented with an amount of resources that is linear in the size of the system, allowing for the resolution of the inverse problem of Eq.~(\ref{main_eq}) also for a large number of spins.

Let us consider the $p$-spin Hamiltonian in transverse field
\be
\hat H_\text{$p$-spin}(\lambda(t)) = -\Gamma (1-\lambda(t))\hat \Sigma_x -\lambda(t)\frac{J}{n^{p-1}}(\hat\Sigma_z)^p\label{pspin0}
\ee
where $\lambda(t)\in[0,1]$ is a time-dependent parameter and
\be
\hat \Sigma_\mu=\nonumber \sum_i\hat\sigma_{i,\mu}.
\ee
We set $J=\Gamma=1$ and $p=3$. For this choice of $p$, the system exhibits a first-order phase transition and the ground state is non-degenerate~\cite{Passarelli4}. The non-degenerate ground state $\hat \rho(t)$ of $\hat H_\text{$p$-spin}(\lambda(t))$ is the state for which we solve the inverse problem.

The model is nonlocal since the interactions in the term $(\hat \Sigma_z)^p$ connect sites regardless of their mutual distance in the spin chain. We look for an optimal parent Hamiltonian for $\hat\rho(t)$ that involves nonlocal interactions connecting a finite number of sites. The maximum number of sites connected by the elementary interactions is called \emph{weight}. We look for a parent Hamiltonian spanned by all the interactions of weight one, two and three. Because of the symmetries inherited by the state $\hat\rho(t)$, these sets must contain only interactions that are invariant for an arbitrary permutation of the chain sites and are the product of an odd number of $\hat \Sigma_y$ operators. Therefore we consider the following sets of elementary interactions
\ba
B_1&=&\{\hat \Sigma_y\},\nonumber\\
B_2&=&B_1\cup\{\hat \Sigma_{xy},\hat \Sigma_{yz}\},\nonumber\\
B_3&=&B_2\cup\{\hat\Sigma_{yyy},\hat\Sigma_{xyx},\hat\Sigma_{zyz},\hat\Sigma_{xxy},\hat\Sigma_{yzz},\hat\Gamma_{xxy},\hat\Gamma_{yzz},\nonumber\\&&\hat\Sigma_{xyz},\hat\Sigma_{xzy},\hat\Sigma_{yxz},\hat\Gamma_{xyz},\hat\Gamma_{xzy},\hat\Gamma_{yxz}\},\nonumber
\ea
where $\hat \Sigma_{\mu_1,..,\mu_n}=(\hat\Sigma_{\mu_1}\dots\hat\Sigma_{\mu_n}+\text{h.\,c.})/2$ and $\hat \Gamma_{\mu_1,..,\mu_n}=i(\hat\Sigma_{\mu_1}\dots\hat\Sigma_{\mu_n}-\text{h.\,c.})/2$.

\begin{figure*}[htp]
\centering
\includegraphics[scale=.35]{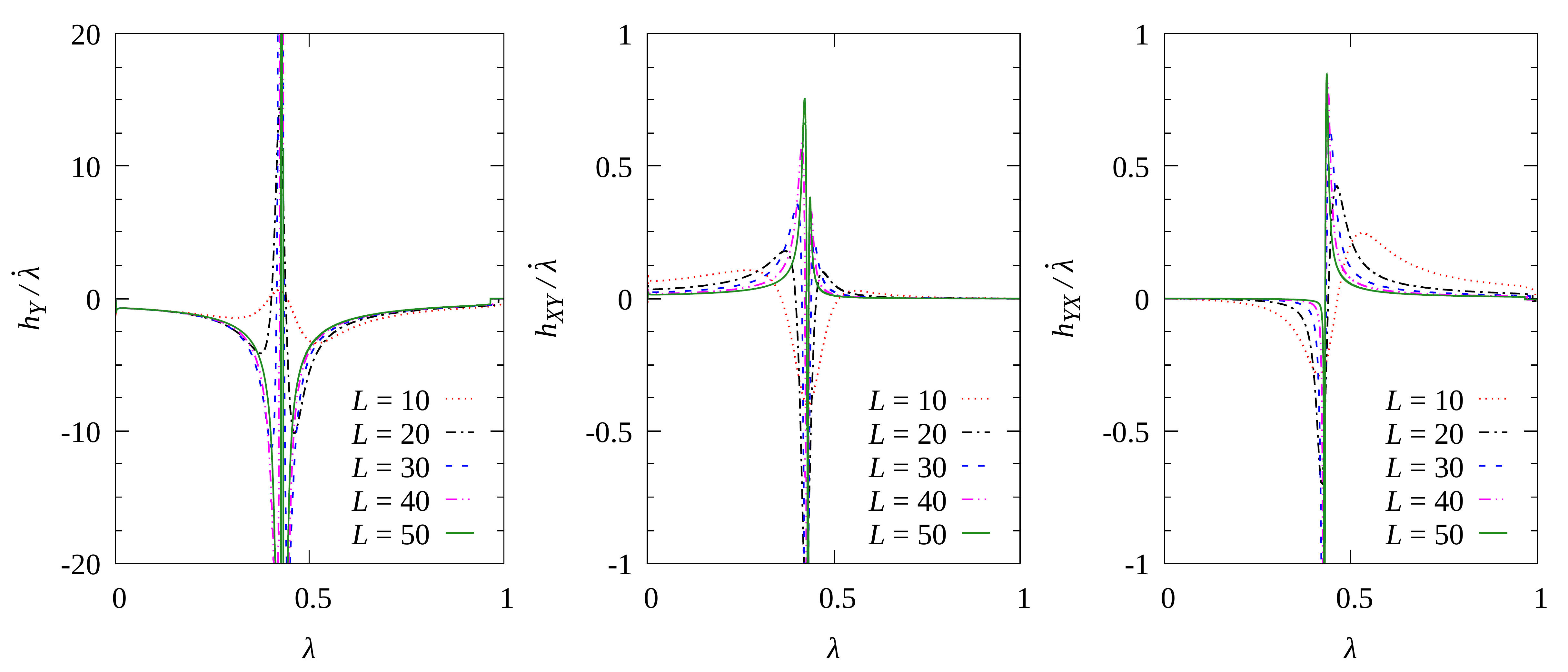}
\caption{Evolution of the (rescaled) optimal coupling $h_i/\dot\lambda$ for different system sizes when the available interactions have maximum weight $w=2$. The couplings shown in figure respectively correspond to the interactions $\hat \Sigma_y$, $\hat \Sigma_{xy}$ and $\hat \Sigma_{yz}$. $\lambda\in[0,1]$ is an arbitrary smooth function of time.}
\label{pspin_couplings_w2}
\end{figure*}

We calculate the couplings of an optimal parent Hamiltonian that is the span of each of the previous sets of interactions by numerically solving Eq.~(\ref{main_eq}). In Figure~\ref{pspin_couplings_w2}, these coupling are shown for different system sizes when the chosen interactions have weight $w=2$. We observe that, in correspondence of noncritical states, the optimal coupling of each interaction converges in the thermodynamic limit, facilitating a scalable implementation of the optimal parent Hamiltonian. Instead, in correspondence of the critical state, the magnitude of the coupling exhibits a large peak whose height increases with the system size. Analogous features are shared by the optimal coupling for weight $w=1$ and $w=2$.

\begin{figure}[htp]
\centering
\includegraphics[scale=.3]{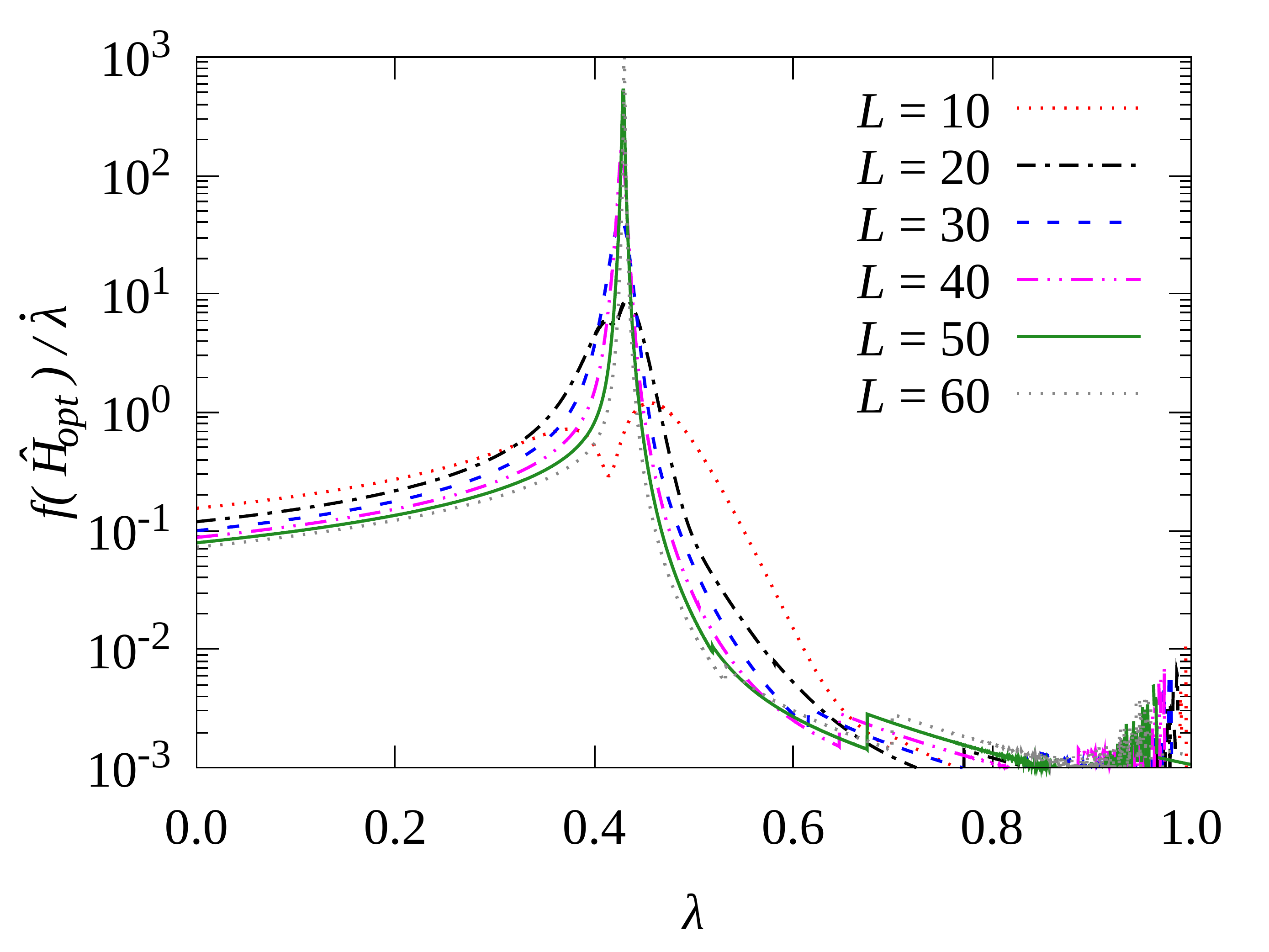}
\caption{Evolution of the (rescaled) optimal local cost $f(\hat H_\text{opt})/\dot\lambda$ for different system sizes when the available interactions have maximum weight $w=2$. $\lambda\in[0,1]$ is an arbitrary smooth function of time.}
\label{pspin_cost_w2}
\end{figure}

\begin{figure}[htp]
\centering
\includegraphics[scale=.32]{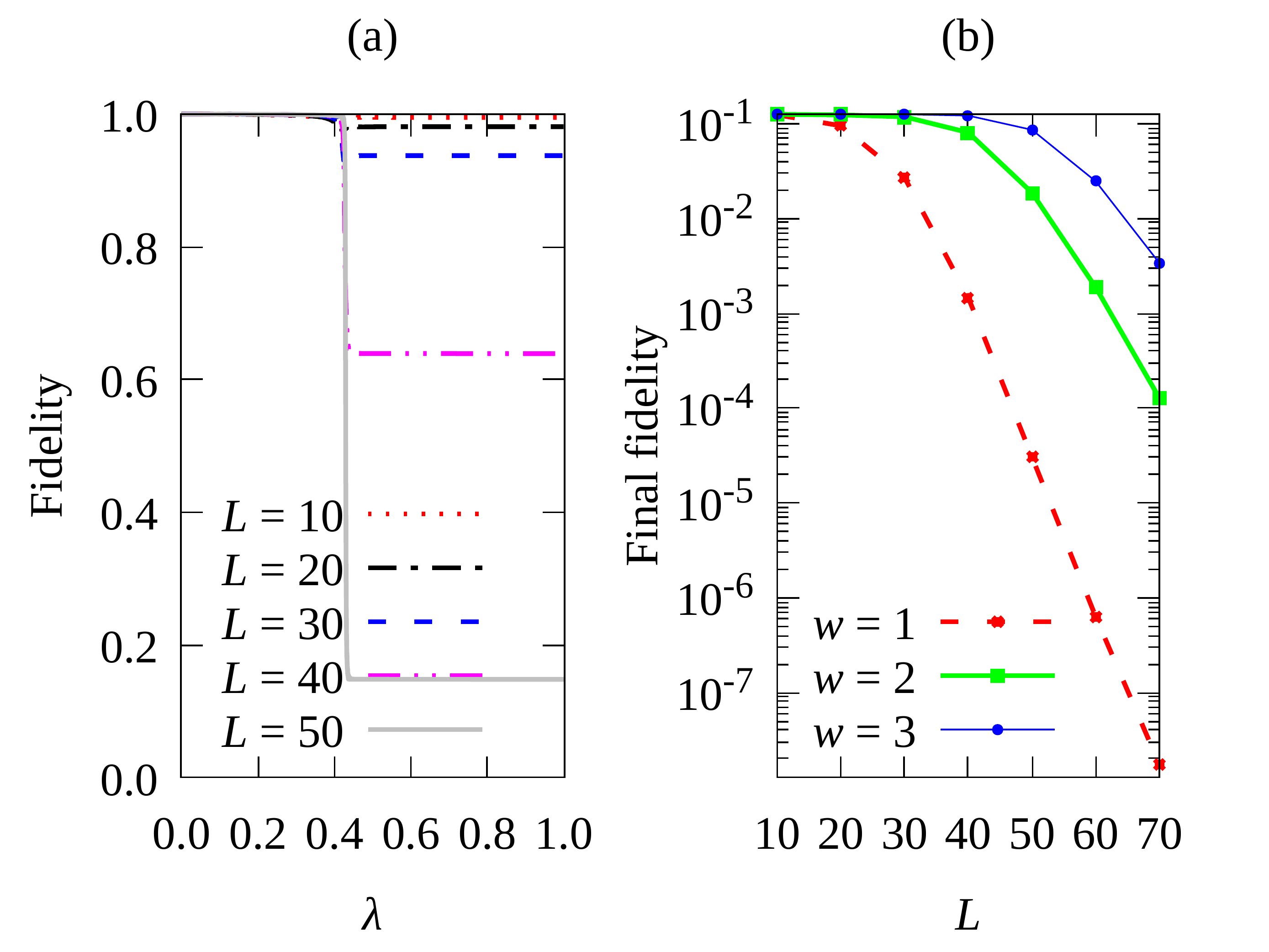}
\caption{In panel (a) evolution of the fidelity between the target state and the state evolved through the optimal parent Hamiltonian of weight $w=2$ for different system sizes, where $\lambda\in[0,1]$ is an arbitrary smooth function of time. In panel (b), fidelity between the target state and the state generated by the optimal parent Hamiltonian at time $T$ as a function of the system size and for different weights of the available interactions.}
\label{pspin_fid_w2}
\end{figure}

The local cost associated to the optimal solution $f(\hat H_\text{opt},t)$ of weight $w=2$ is shown in Figure~\ref{pspin_cost_w2}. Here, we can observe that the cost decreases and converges to a minimum value for noncritical states when the system size is increased: a better parent Hamiltonian for a $p$-spin model ground state exists when the size of the system is enlarged. As we can see in Figure~\ref{pspin_fid_w2} (a), this feature is inherited by the fidelity between the target state and the driven one: larger $p$-spin noncritical states are easier to be driven. The situation is reversed in correspondence of the critical states, where the cost of the optimal solution drastically increases with the system size, generating a sharp drop of the fidelity in Figure~\ref{pspin_fid_w2}: as in the Ising model, the optimal parent Hamiltonian fails to generate a quantum phase transition. Analogous features are shared by the cost function and the fidelity for weight $w=1$ and weight $w=3$.

The final fidelity between the target state and the driven one for $\lambda(t)=1$ is shown in Figure~\ref{pspin_fid_w2} (b) for different sizes of the spin chain and for different weights of the interactions. As expected, the fidelity is drastically improved using operators of larger weights, since the basis $B_i$, and therefore the space $TL_{\hat\rho}$ on which the exact evolution is projected, is expanded.

As anticipated, the search for an optimal Hamiltonian capable of generating the path of ground states for the $p$-spin model in exam has been largely investigated in the recent literature in the context of shortcuts to adiabaticity. In these works, the authors look for an optimal countediabatic Hamiltonian by minimizing the cost functional proposed by Sels and Polkovnikov~\cite{Polkovnikov}. The differences between the results of these papers and our results can be easily understood by investigating the difference between the minimization of Eq.~(\ref{cost_functional}) and the counterdiabatic cost proposed in Ref.~\onlinecite{Polkovnikov}. Section~\ref{optimal-driving-ground} of this paper is devoted to this aim.

\subsection{Linear interpolation of ground states}

In the previous examples, we have solved the inverse problem for a path of states that have been explicitly built as ground states of a given time-dependent  Hamiltonian. However, our method is suited to the search of an optimal parent Hamiltonian for a generic time-dependent state. Here we attack the inverse problem for a time-dependent linear interpolation of a pair of states. The initial and final states are ground states of the $p$-spin Hamiltonian $\hat H_\text{$p$-spin}(\lambda)$ at $\lambda=0$ and $\lambda=1$. In this way, we can compare the performances of our method for different paths of states with the same extremal points.

The target time-dependent state is
\be
\ket{\psi(t)}=Z(t)\left[\cos\left(2\pi\lambda(t)\right)\ket{\psi_0}+\sin\left(2\pi\lambda(t)\right)\ket{\psi_1}\right]\nonumber
\ee
where $\ket{\psi_0}$ and $\ket{\psi_1}$  are the ground states of the Hamiltonian in Eq.~\eqref{pspin0} with $\lambda=0,1$, respectively, and $Z(t)$ is a normalization factor such that $\lVert\ket{\psi(t)}\rVert=1$ at each time $t$. $\lambda(t)$ is an arbitrary smooth function of time such that $\lambda(0)=0$ and $\lambda(T)=1$. We represent this state through the density matrix $\hat\rho(t)=\ket{\psi(t)}\bra{\psi(t)}$. We construct an optimal parent Hamiltonian with the operators of weight one, two and three  of the sets $B_1$, $B_2$ and $B_3$.

\begin{figure*}[htp]
\centering
\includegraphics[scale=.35]{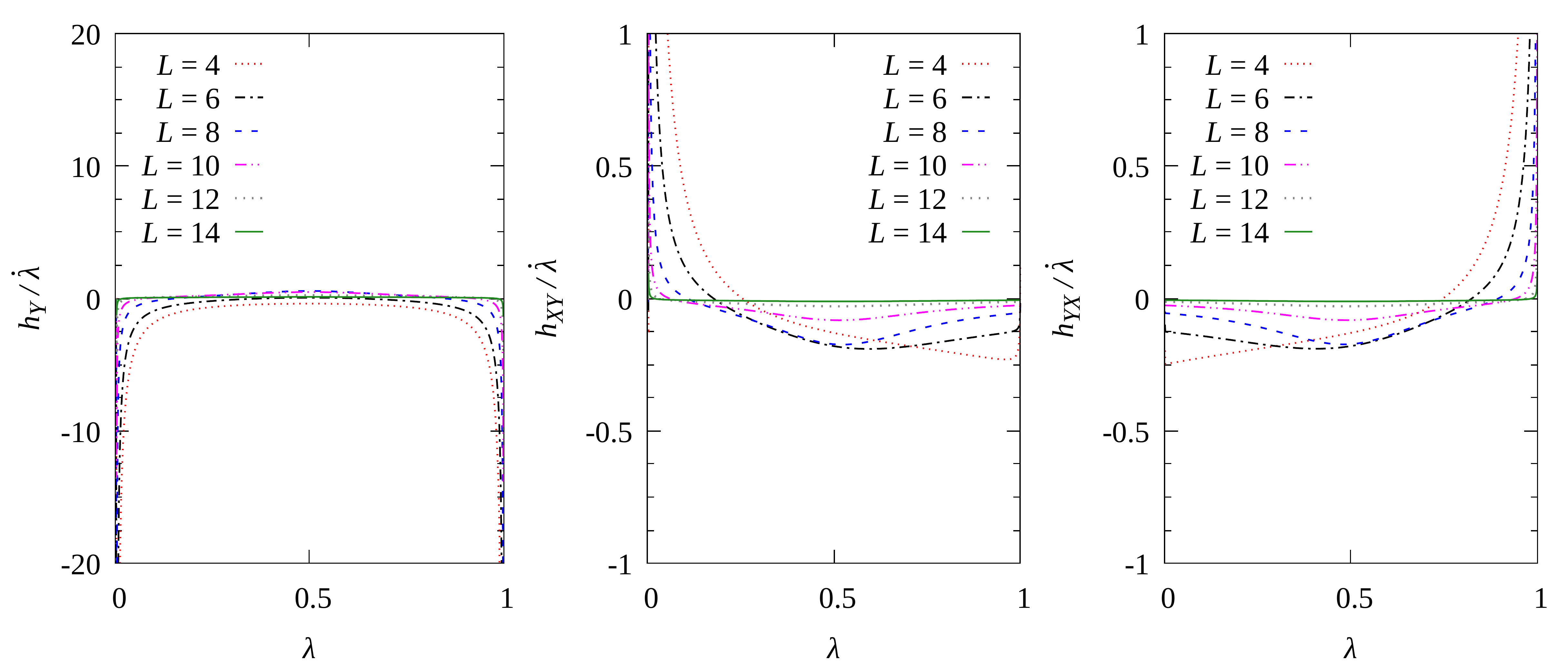}
\caption{Evolution of the (rescaled) optimal coupling $h_i/\dot\lambda$ for different system sizes when the available interactions have maximum weight $w=2$. $\lambda\in[0,1]$ is an arbitrary smooth function of time. The couplings shown in figure respectively correspond to the interactions $\hat \Sigma_y$, $\hat \Sigma_{xy}$ and $\hat \Sigma_{yz}$.}
\label{pspin2_couplings_w2}
\end{figure*}

\begin{figure}[htp]
\centering
\includegraphics[scale=.3]{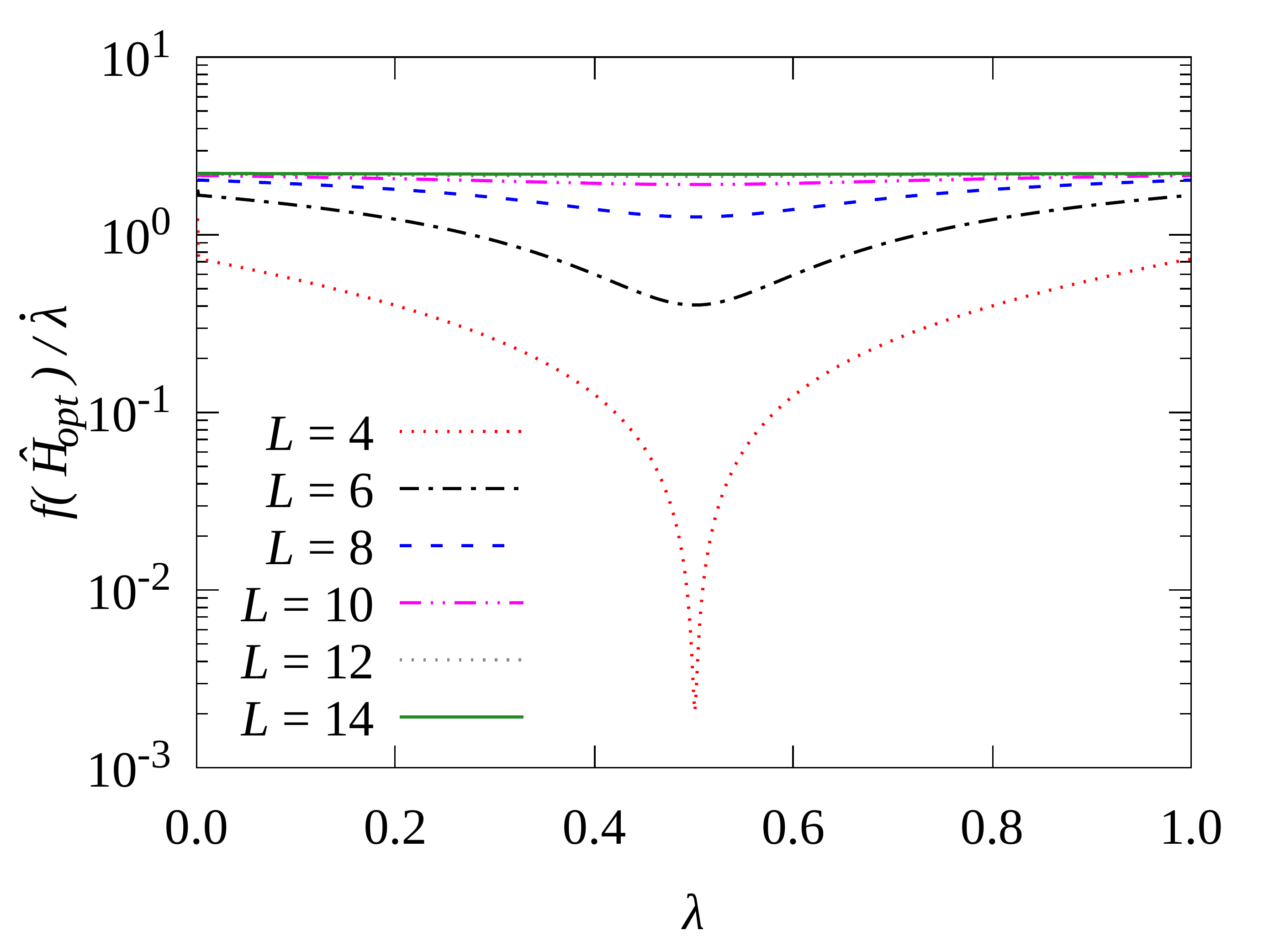}
\caption{Evolution of the (rescaled) optimal local cost $f(\hat H_\text{opt})/\dot\lambda$ for different system sizes when the available interactions have maximum weight $w=2$. $\lambda\in[0,1]$ is an arbitrary smooth function of time.}
\label{pspin2_cost_w2}
\end{figure}

The solution to Eq.~(\ref{main_eq}) in the space spanned by the operators of weight two leads to the optimal couplings shown in Figure~\ref{pspin2_couplings_w2}. Here, we can observe that the magnitude of the coupling decreases with the system size and is maximized at the initial and final times. A similar behavior is shared by the local cost associated to the optimal Hamiltonian, which is shown in Figure~\ref{pspin2_cost_w2} for different numbers of spins. We can see that the local cost increases with the system size and is maximized at the initial and final times. The behavior of the couplings and of the total cost at the extremal times seems to reflect the fact that constructing an interpolation of ground states with the available interactions is very hard.

\begin{figure}[htp]
\centering
\includegraphics[scale=.32]{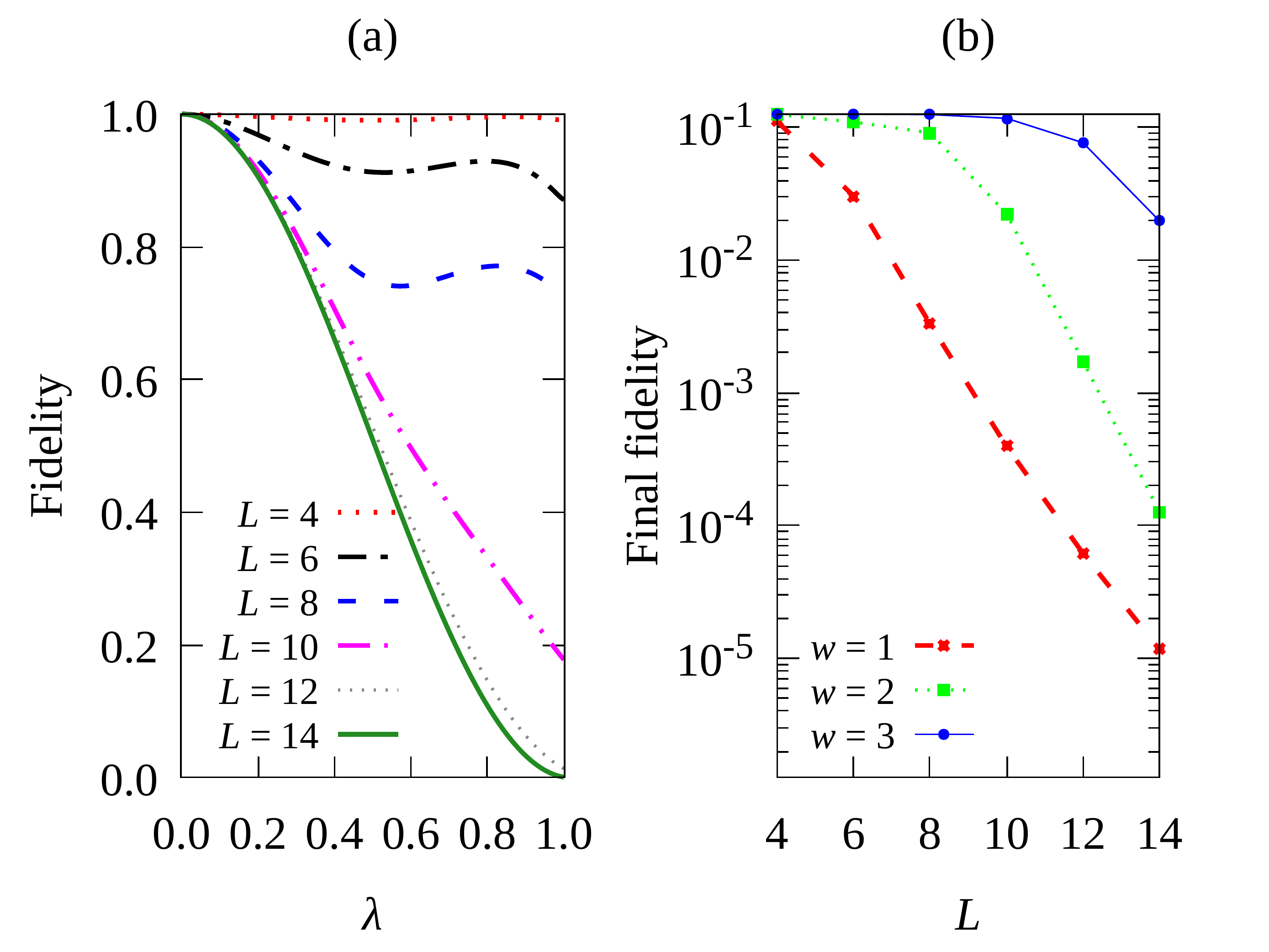}
\caption{Panel (a): evolution of the fidelity between the target state and the state evolved through the optimal parent Hamiltonian of weight $w=2$ for different system sizes, where $\lambda\in[0,1]$ is an arbitrary smooth function of the time. Panel (b): fidelity between the target state and the state generated by the optimal parent Hamiltonian at time $T$ for different sizes of the system and for different weights of the available interactions.}
\label{pspin2_fid_w2}
\end{figure}

The behavior of the optimal cost function is reflected in the fidelity between the target state and the optimally driven one in Figure~\ref{pspin2_fid_w2} (a), where larger drops in fidelity correspond to larger values of the local cost function.

The final fidelity between the target state and the driven one at time $T$ is shown in Figure~\ref{pspin2_fid_w2} (b) for different system sizes and different weights of the allowed interactions. As expected, we can see that a larger weight corresponds to a larger final fidelity.

Both the time-dependent state analyzed here and the one analyzed previously start from and end at the same points, but, for fixed number of spins, the cost associated to the time-dependent state studied here is larger and the final fidelity is dramatically lower. We can state that the path of states of the previous section is more accessible with the interactions in exam. The investigation of the concept of  ``accessibility" and the search for a more accessible path of states are stimulating questions but are beyond the aims of this work. Here, driven by the observation that building interpolations of states seems more difficult than building ground states, we just conjecture a direct relationship between the evolution of the entanglement and the accessibility of the time-dependent state.

\section{OPTIMAL DRIVING AND SHORTCUT TO ADIABATICITY}

\label{optimal-driving-ground}
Two examples from the previous section have been devoted to the search for an optimal parent Hamiltonian for the time-dependent ground state $\text{GS}(\lambda(t))$ of an Hamiltonian $\hat H_\text{a}(\lambda(t))$. In this context, the inverse problem is exploited to perform a shortcut to adiabaticity. A shortcut to adiabaticity is a protocol to drive a quantum state to the ground state of a target Hamiltonian in a shorter time than that required by an adiabatic process~\cite{shortcuts1,shortcuts2,shortcuts3,shortcuts4,shortcuts5}. To reach this goal one needs to suppress Landau-Zener transitions to the excited eigenstates, which naturally arise when the control parameters of the system rapidly change. One of the most exploited shortcuts to adiabaticity consists in suppressing these transition through the so-called counterdiabatic potential associated to $\hat H_\text{a}(\lambda(t))$\cite{Berry}. This potential also is one of the exact parent Hamiltonians for $\text{GS}(\lambda(t))$. However, since the counterdiabatic potential is the span of interactions that are very hard to engineer in an experiment, one looks for an approximate potential that only involves a given set of interactions $B_L=\{L_i\}$. Sels and Polkovnikov~\cite{Polkovnikov} proposed a local approximation of the counterdiabatic potential through the minimization of a cost functional depending on the adiabatic Hamiltonian. This local counterdiabatic potential can be exploited to optimally control a time-dependent state, so we need to understand which is the relationship between the local counterdiabatic potential and our optimal solution to the inverse problem. This section is devoted to the investigation of this analogy.

There are two fundamental differences between the counterdiabatic potential and the application of the exact parent Hamiltonian as a shortcut to adiabaticity. First of all, in order to construct a counterdiabatic potential, one needs to know the adiabatic Hamiltonian, but the reconstruction of the latter from a time-dependent state $\hat \rho(t)$ is not a trivial problem and could be considered a particular solution to the optimal inverse problem.
Moreover, the optimal counterdiabatic Hamiltonian is suited to the simultaneous driving of all the eigenvalues of the adiabatic Hamiltonian. This can be easily shown if one consider that the  potential in exam is the minimum of the functional $S_{\hat H_\text{a}}(\hat A^*)=\int_0^T \Tr\left[(\partial_t \hat H_\text{a}+i[\hat A^*,\hat H_\text{a}])^2\right]dt$ with respect to the candidate counterdiabatic potential $\hat A^*$, where $\hat H_\text{a}(t)$ is the adiabatic Hamiltonian. One can show (see Appendix~\ref{appendix_polk}) that the minimum of $S_{\hat H_\text{a}}$ is also the minimum of
\be
S_{\hat H_\text{a}}'(\hat A^*)=\int_0^T \lVert\sum_i E_i (\partial_t\hat\rho_i+i[\hat A^*,\hat\rho_i])\rVert dt,
\ee
where the $E_i$ and the $\hat\rho_i$ are the eigenvalues and the eigenvectors of $\hat H_\text{a}$ respectively. This last functional is related to our cost functionals $\mathcal{F}_{\hat\rho_i}$ associated to the inverse problems for the eigenstates $\hat \rho_i(t)$ as follows:
\be
S_{\hat H_\text{a}}'\leq\sum_i\lvert E_i\rvert \mathcal{F}_{\hat\rho_i}
\ee
Therefore, the minimum of $S_{\hat H_\text{a}}$ optimally drives all the eigenstates of the adiabatic Hamiltonian, and the driving of each of these states $\hat\rho_i$ is obtained by minimizing the inverse problem cost functional $\mathcal{F}_{\hat\rho_i}$ and the corresponding energy has the role of a penalty. As a consequence, a minimum for the potential $S_{\hat H_\text{a}}$ is generally different from a minimum of $\mathcal{F}$. As an example, let us consider the transformation $\hat H_\text{a}\rightarrow \hat H_\text{a}+d\hat H_\text{a}$ for the Hamiltonian $\hat H_\text{a}$, where $\hat H_\text{a}=\text{diag}(E_1,E_2,E_3)$ and
\be
d\hat H_\text{a}=dt\dot\lambda \begin{pmatrix}
	0 & (E_2-E_1) & 0\\
	(E_1-E_2) & 0 & 2(E_3-E_2)\\
	0 & 2(E_2-E_3) & 0\\
\end{pmatrix}i,\nonumber
\ee
in which $i$ is the imaginary unit. If we look for the minimum of the cost function $S_{\hat H_\text{a}}$ in the space spanned by the operator
\be
\hat P=\begin{pmatrix}
	0 & 1 & 0\\
	1 & 0 & 1\\
	0 & 1 & 0\\
\end{pmatrix},\nonumber
\ee
we obtain the counterdiabatic potential
\[\hat H_\text {CD}=-\dot\lambda\frac{(E_1-E_2)^2+2(E_3-E_2)^2}{(E_1-E_2)^2+(E_3-E_2)^2}\hat P.\]
The optimal parent Hamiltonian for the ground state of $\hat H_\text{a}$ instead corresponds to the minimum of the associated cost functional $\mathcal{F}$. This can be obtained by replacing $E_2$ and $E_3$ with zero in the last expression. Now the minimization leads to
\[\hat H_\text {opt}=-\dot\lambda \hat P.\]
The difference between these Hamiltonians mirrors the different goals for which the corresponding potential are designed: while the first one is suited to simultaneously drive all the eigenstates of $\hat H_\text{a}$, the second one exploits all the available resources to drive only its ground state.

\section{CONCLUSIONS AND OUTLOOK}
\label{conclusions}

In this paper, we have illustrated the exact solutions to the quantum time-dependent inverse problem. When the space of Hamiltonians is restricted to a subset of realistic interactions, we have also defined an engineerable optimal solution to this problem. Beyond the reconstruction of the time-dependent parent Hamiltonian, an optimal solution with a sufficiently low cost also drives the initial state along the target evolution with a high fidelity. As a consequence, the applications of this methods include both the manipulation of quantum states and the investigation of the interactions affecting the system. Exploiting the geometrical interpretation of the cost functional, we have found the analytical expression of the optimal Hamiltonian in terms of the state and its derivative at each time. Finally, we have demonstrated the performance of our method in driving a time-dependent ground state of the Ising model in transverse field and of the $p$-spin model.

Beyond the direct extension to time-dependent quantum verification and Hamiltonian learning, an interesting application of our method consists in calculating the correlations in Eq.~(\ref{main_eq}) using a quantum device. In this way, one could use our equation to look for a simplified version of the driving protocol that the device exploits to implement the state $\hat\rho(t)$.

In this paper, we have focused our attention to a particular optimal solution to the inverse problem, whose performance in driving the state does not depend on the total time of the process. A different approach consists in selecting the adiabatic solution to the inverse problem, defined as the Hamiltonians capable of generating the driving with high fidelity for a sufficiently slow evolution. Such an adiabatic solution has to simultaneously satisfy two constraints: it has to be a symmetry for the state at any time, and the state must belong to a nondegenerate eigenspace of the Hamiltonian at any time. As previously shown in the literature, local symmetries can be easily found as the kernel of the matrix $V$, but until now there is no systematic way to select the local Hamiltonians that also obey the second constraint. The design of a cost functional capable of this selection is one of the most interesting natural prosecution of this work.

In Section~\ref{optimal-ham}, we have also defined a geometrical measure of the difficulty of generating a target time-dependent state exploiting the local interaction, that is the \emph{accessibility angle}. This geometrical notion could be used to select the best path to connect quantum states by looking for geodesics of a related metric. Moreover, the physical meaning of this object has a strong analogy with the idea of a geometry of quantum complexity\cite{Nielsen1,Nielsen2,susskindcomplexity}. This insight could be consolidated by showing a functional relation between the cost associated to the optimal parent Hamiltonian and the complexity of the best locally driven approximation of $\hat\rho(t)$. Once this relation has been proven, the optimal cost function can be associated to an \emph{accessibility metric} on the manifold of quantum states, and its relation to the complexity metric proposed by Nielsen\cite{Nielsen1,Nielsen2,susskindcomplexity} could be clarified.

Finally, the central role of the density matrix in our approach suggests the possibility of an extension to open quantum systems and the better performance demonstrated in driving states with exponential decaying correlations is a good reason to support the application to Matrix Product States.

\section*{ACKNOWLEDGEMENTS}

P. L. acknowledges Università di Napoli Federico II financial support under the initiative \emph{Programma di scambi internazionali per la mobilita' di breve durata} D.R. n. 2243 del 12/06/2017.

R. F. acknowledges partial financial support from the Google Quantum Research Award.

R. F. research has been conducted within the framework of the Trieste Institute for Theoretical Quantum Technologies (TQT).


\appendix

\section{Relationship between cost and fidelity}
\label{appendix-bound}

In this section, we prove the inequality in Eq.~\eqref{fidelity_bound} of Section~\ref{optimal-ham}.
Given a trial Hamiltonian $\hat H(t)$, the associated cost functional is
\begin{equation*}
	\mathcal{F}[\hat H]=\int_0^Tdt f(\hat H,t),
\end{equation*}
where
\begin{equation*}
	f(\hat H,t)=\lVert\partial_t \hat \rho(t)+i[\hat H(t),\hat \rho(t)]\rVert.
\end{equation*}

Here we show that, at the final time $T$, the Hilbert-Schmidt distance between the target state $\hat \rho(T)$ and the state
\begin{equation*}
	\hat \rho_\text{opt}(t) =\left(\mathcal{T}e^{-i\int_0^t\hat H(t')dt'}\right)\hat\rho(0)\left(\mathcal{T}e^{-i\int_0^t\hat H(t')dt'}\right)^\dag
\end{equation*}
evolved with the trial Hamiltonian, is upper bounded by the cost $\mathcal{F}[\hat H]$.

We start from the inequality
\begin{equation}\label{supp1}
	\lVert\hat\rho(T)-\hat\rho_\text{opt}(T)\rVert \leq \int_0^T dt \, \big\lvert \partial_t {\lVert \hat\rho(t)-\hat\rho_\text{opt}(t)\rVert} \big\rvert.
\end{equation}
The derivative of the squared norm in the last equation is
\begin{align*}
	\partial_t\lVert\hat\rho-\hat\rho_\text{opt}\rVert^2&=2\Tr\bigl[(\hat\rho-\hat\rho_\text{opt})\partial_t(\hat\rho-\hat\rho_\text{opt})\bigr]\\
	&=2\Tr\bigl[(\hat\rho\!-\!\hat\rho_\text{opt})(\partial_t \hat\rho+i[\hat H,\hat\rho]\!-\!i[\hat H,\hat\rho\!-\!\hat\rho_\text{opt}])],
\end{align*}
where we have exploited the Leibniz rule and we have written $\partial_t\hat\rho_\text{opt}$ as $-i[\hat H,\hat\rho_\text{opt}]$. Taking into account the Hilbert-Schmidt orthogonality between $\hat A$ and $[\hat A,\hat B]$ for any pair of operators $ \hat A $ and $ \hat B $, the latter becomes
\begin{equation*}
	\partial_t\lVert\hat\rho-\hat\rho_\text{opt}\rVert^2=2\Tr\bigl[(\hat\rho-\hat\rho_\text{opt})(\partial_t \hat\rho+i[\hat H,\hat\rho])\bigr].
\end{equation*}
Finally, we apply the Cauchy-Schwartz inequality to obtain
\begin{equation*}
	\big\lvert\partial_t{\lVert\hat\rho-\hat\rho_\text{opt}\rVert}^2\big\lvert\leq 2\lVert\hat\rho-\hat\rho_\text{opt}\rVert \, f(\hat H,t).
\end{equation*}
On the other hand, the Leibniz rule also holds in the following form:
\begin{equation*}
	\partial_t{\lVert\hat\rho-\hat\rho_\text{opt}\rVert}^2= 2\lVert\hat\rho-\hat\rho_\text{opt}\rVert\,\partial_t\lVert\hat\rho-\hat\rho_\text{opt}\rVert,
\end{equation*}
hence
\begin{equation*}
	\big\lvert\partial_t\lVert\hat\rho-\hat\rho_\text{opt}\rVert\big\rvert\leq f(\hat H,t).
\end{equation*}
Replacing the latter in Eq.~\eqref{supp1}, we obtain the upper bound
\begin{equation}
	\lVert\hat\rho(T)-\hat\rho_\text{opt}(T)\rVert\leq\int_0^T dt \, f(\hat H,t) =\mathcal{F}[\hat H].
\end{equation}
For pure states, we have $ \lVert\hat\rho(T)-\hat\rho_\text{opt}(T)\rVert=2(1-F(\hat\rho(T),\hat\rho_\text{opt}(T)))$, where $F$ is the squared fidelity, therefore the latter equation corresponds to the inequality in Eq.~(\ref{fidelity_bound}).

\section{Fubini-Study and Hilbert-Schmidt distance}
\label{appendix-fs}

Here we demonstrate the relationship between the Hilbert-Schmidt metric for pure states and the Fubini-Study metric. Let $\{\hat T_n\}$ be a set of Hermitian generators for a basis $\{\ket{\partial_n\psi}\}$ of a tangent space for $\ket{\psi}$, that is, $\ket{\partial_n\psi}=i\,\hat T_n \ket{\psi}$ for each $\ket{\partial_n\psi}$. The derivative of a vector in an arbitrary direction can be written as
\begin{equation*}
\ket{d\psi}=i\,\delta^n \ket{\partial_n\psi}=i \,\delta^n \hat T_n\ket{\psi}
\end{equation*}
where the $\delta^n$ are coordinates. The Fubini-Study metric is defined as
\begin{equation*}
	g(d\psi, d\psi) = \text{Re}(\langle d\psi|d\psi\rangle-\langle \psi|d\psi\rangle\langle d\psi|\psi\rangle).
\end{equation*}

In the basis defined above, the coordinate representation of the Fubini-Study metric is the matrix $g_{ij}$ such that
\begin{equation*}
	g(d\psi,d\psi)=\delta^n\delta^mg_{nm}.
\end{equation*}

We show that this matrix is proportional to the correlation matrix $V_{nm}$ of the operators $\{\hat T_n\}$:
\begin{align*}
	\delta^n\delta^m g_{nm}&=\text{Re}(\langle d\psi|d\psi\rangle-\langle \psi|d\psi\rangle\langle d\psi|\psi\rangle)\\
	&= \delta^n\delta^m \text{Re}(\langle \psi|\hat T_n\hat T_m|\psi\rangle-\langle \psi|\hat T_n|\psi\rangle\langle \psi|\hat T_m|\psi\rangle)\\
	&= \frac{\delta^n\delta^m}{2} \text{Re}\left(\langle \psi|\{\hat T_n,\hat T_m\}|\psi\rangle-\langle \psi|\hat T_n|\psi\rangle\langle \psi|\hat T_m|\psi\rangle\right)\\
	&= \frac{1}{2}\delta^n\delta^m V_{nm}.
\end{align*}

The correlation matrix $V_{nm}$ is also the metric associated to the Hilbert-Schmidt length of tangent vectors:
\begin{align*}
	{\lVert d \psi\rVert}^2 &= \Tr\bigl[(\ket{d\psi}\bra{\psi}+ \text{h.\,c.})^2\bigr]\\
	&= \Tr\bigl[(i\, \delta^n[\hat T_n,\hat \rho])^2\bigr]\\
	&=-\delta^n\delta^m \Tr\bigl[(\hat T_n\hat \rho-\hat \rho \hat T_n)(\hat T_m\hat \rho\!-\!\hat \rho \hat T_m)\bigr]\\
	&=\delta^n\delta^mV_{nm}.
\end{align*}
Therefore, the Hilbert-Schmidt length of tangent vectors is twice the Fubini-Study length, and the correlation matrix associated to the operators $\hat T_n$ is twice the Fubini-Study metric on the tangent space with basis $\{i\,\hat T_n\ket{\psi}\}$.

\section{Ground states of the Ising model}
\label{appendix-Ising}

Here we show some details of the reconstruction of the exact and optimal parent Hamiltonians for the ground states of the Ising model that have been omitted from the main text.

\subsection{Pseudo-spins representation of the Hamiltonian}
The Hamiltonian of the Ising model is
\begin{equation*}
\hat H_\text{I}(\lambda)=-\sum_{i=1}^N\hat\sigma_{i,x}\hat\sigma_{i+1,x}-\lambda\sum_{i=1}^N \hat\sigma_{i,z},
\end{equation*}
where, for simplicity, we suppose that the number of spins $N$ is even and that there are periodic boundary conditions $\hat\sigma_{N+1,\mu}=\hat\sigma_{1,\mu}$.

Exploiting the Jordan-Wigner transformation~\cite{jordan_wigner}
\begin{align*}
	\hat K_j&=\prod_1^{j-1}(\hat\Id-2\hat c_j^\dag \hat c_j)\\
	\hat\sigma_{j,z}&=(\hat\Id-2\hat c_j^\dag \hat c_j)\\
	\hat\sigma_{j,x}&=\hat K_j(\hat c_j^\dag+\hat c_j)\\
	\hat\sigma_{j,y}&=i\,\hat K_j(\hat c_j^\dag-\hat c_j),
\end{align*}
$\hat H_\text{I}$ can be written as
\begin{align*}
	\hat H_\text{I}&=-\sum_1^{L-1}\left(\hat c_j^\dag \hat c_{j+1}^\dag+\hat c_{j}^\dag \hat c_{j+1}+\text{h.\,c.}\right)+\lambda\sum_{j=1}^L(2\hat c_j^\dag \hat c_j-\hat\Id)\nonumber\\
	&+\left(\hat c_L^\dag \hat c_{L+1}^\dag+\hat c_{L}^\dag \hat c_{L+1}+\text{h.\,c.}\right)\hat Q,
\end{align*}
where we have defined the parity operator $\hat Q=(-\hat \Id)^{\sum_j \hat c_j^\dag \hat c_j}$. This operator is a symmetry for the Hamiltonian and divides the Hilbert space in two parity sectors. The ground state of the Ising model lies in the even sector where $ Q=1$. Therefore, from now, with a little abuse of notation we represent the Hamiltonian only in this sector. Once defined the Fourier's operators
\begin{align*}
	\hat c_k&=\frac{e^{-i\pi/4}}{\sqrt{L}}\sum_{j=1}^Le^{-ikj}\hat c_j,\\
	\hat c_j&=\frac{e^{i\pi/4}}{\sqrt{L}}\sum_{k\in\mathcal{K}}e^{ikj}\hat c_k,
\end{align*}
in the even parity sector the Hamiltonian is
\begin{align*}
	\hat H_\text{I}&=\sum_{k\in\mathcal{K}^{+}}\hat H_k\\
	\hat H_k&=\epsilon_k\bigl[\cos(\theta_k)(\hat c_k^\dag \hat c_k-\hat c_{-k}\hat c_{-k}^\dag)\\&+\sin(\theta_k)(\hat c_k^\dag \hat c_{-k}^\dag+\hat c_{-k}\hat c_k)\bigr],
\end{align*}
with
\begin{align*}
	\epsilon_k(\lambda)&=2\sqrt{(h-J\cos(k))^2+(J\sin(k))^2},\\
	\theta_k(\lambda)&=-\arctan\left(\frac{J\sin(k)}{h-J\cos(k)}\right),
\end{align*}
and $\mathcal{K}=\{\pm\frac{(2n-1)\pi}{L}\text{, with }n=1,...,L/2\}$.

Finally, we can introduce the pseudo-spins operators
\begin{align*}
	\tilde\sigma_k^x&=\hat c_k^\dag \hat c_{-k}^\dag+\hat c_{-k}\hat c_k\\
	\tilde\sigma_k^y&=-i(\hat c_k^\dag \hat c_{-k}^\dag-\hat c_{-k}\hat c_k)\\
	\tilde\sigma_k^z&=\hat c_k^\dag \hat c_k-\hat c_{-k}\hat c_{-k}^\dag
\end{align*}
and write the Hamiltonians $\hat H_k$ as
\begin{equation*}
	\hat H_k=\epsilon_k\left[\cos(\theta_k)\tilde\sigma_k^z+\sin(\theta_k)\tilde\sigma_k^x\right].
\end{equation*}

Since the operators $\tilde\sigma_k^\mu$ obey spin commutation rules, in this representation the Ising Hamiltonian is a Hamiltonian of noninteracting spins in an inhomogeneous transverse field, with trivial eigenvalues and eigenvectors.

The pseudo-spin operators can be represented as a function of the real-space spins by inverting all the previous transformations as follows:
\begin{align*}
	\tilde\sigma_k^y&=-i(\hat c_k^\dag \hat c_{-k}^\dag-\hat c_{-k}\hat c_k)\\
	&=\frac{1}{L}\sum_{j,j'}e^{ik(j-j')}(\hat c_j^\dag \hat c_{j'}^\dag+\hat c_j\hat c_{j'})\\
	&=-\frac{i}{L}\sum_{j,j'}\sin[k(j'-j)](\hat c_j^\dag \hat c_{j'}^\dag+\hat c_j\hat c_{j'})\\
	&=\frac{1}{L}\sum_{j'>j}\sin[k(j'-j)]\hat{\mathcal{S}}_{jj'},
\end{align*}
where we have defined the spin string operators
\ba
	\hat{\mathcal{S}}_{jj'}&=&\hat\sigma_{j,x}\hat\sigma_{j+1,z}\dots\hat\sigma_{j'-1,z}\hat\sigma_{j',y}\nonumber\\&+&\hat\sigma_{j,y}\hat\sigma_{j+1,z}\dots\hat\sigma_{j'-1,z}\hat\sigma_{j',x}.\nonumber
\ea

\subsection{Exploiting symmetries}

We show how to exploit the symmetries of the Ising model to find a simple expression for the optimal parent Hamiltonian. The coefficients of the optimal parent Hamiltonian are solution to Eq.~\eqref{main_eq}. If we replace the time derivative $\partial_t\hat\rho$ with $-i[\hat H_\text{p},\hat\rho]$ where $\hat H_\text{p}$ is the exact parent Hamiltonian, this equation becomes
\begin{equation}\label{supp_2}
	\langle \hat H_\text{p}, \hat L_a\rangle=\sum_bh_b\langle \hat L_b, \hat L_a\rangle,
\end{equation}
where the $\hat L_a\in
B_L=\{\hat\sigma_{i,\mu}\}\cup\{\hat\sigma_{i,\mu} \hat\sigma_{i+1,\nu}\}\nonumber$ are Pauli operators acting on one spin and two adjacent spins and $\langle \hat A, \hat B\rangle=\Tr[\hat\rho (\hat A\hat B+\hat B\hat A)]-2\Tr(\hat\rho \hat A)\Tr(\hat\rho \hat B)$ is a connected correlation.

First of all, we note that the system is symmetric for a reflection of the $x$-axis ($\hat\sigma_{i,x}\rightarrow-\hat\sigma_{i,x}$) or of the $y$-axis ($\hat\sigma_{i,y}\rightarrow-\hat\sigma_{i,y}$). As a consequence, all the expectation values of an odd number of Pauli operators on these axis are null and Eq.~\eqref{supp_2} can be represented in a block-matrix form as follows
\begin{equation*}
\begin{pmatrix}
	\vec{0}\\
	\langle \hat H_\text{p} ,\vec{XY}\rangle\\
	\langle \hat H_\text{p} ,\vec{YX}\rangle\\
\end{pmatrix}=
\begin{pmatrix}
	A & 0 & 0\\
	0 & \langle \vec{XY} ,\vec{XY}\rangle & \langle \vec{YX} ,\vec{XY}\rangle\\
	0 & \langle \vec{XY} ,\vec{YX}\rangle & \langle \vec{YX} ,\vec{YX}\rangle\\
\end{pmatrix}
\begin{pmatrix}
	\vec{h}_0\\
	\vec{h}_{xy}\\
	\vec{h}_{yx}\\
\end{pmatrix},
\end{equation*}
where $\vec{XY}=(\hat\sigma_{1,x}\hat\sigma_{2,y},\hat\sigma_{2,x}\hat\sigma_{3,y},...,\hat\sigma_{L,x}\hat\sigma_{1,y})^T$ and $\vec{YX}=(\hat\sigma_{1,y}\hat\sigma_{2,x},\hat\sigma_{2,y}\hat\sigma_{3,x},...,\hat\sigma_{L,y}\hat\sigma_{1,x})^T$. It follows that the optimal parent Hamiltonian is
\begin{equation*}
	\hat H_\text{opt}=\sum_i\left[(\vec{h}_{xy})_i\hat\sigma_{i,x}\hat\sigma_{i+1,y}+(\vec{h}_{yx})_i\hat\sigma_{i,y}\hat\sigma_{i+1,x}\right].
\end{equation*}
The symmetry of the Ising Hamiltonian for a translation or an inversion of the spin chain, that is, $\hat\sigma_{i,\mu}\rightarrow\hat\sigma_{i+1,\mu}$ and $\hat\sigma_{i,\mu}\rightarrow\hat\sigma_{-i,\mu}$, is inherited by the optimal parent Hamiltonian, which becomes 
\begin{equation}
\hat H_\text{opt}(t)=h(t)\sum_i\left[\hat\sigma_{i,x}\hat\sigma_{i+1,y}+\hat\sigma_{i,y}\hat\sigma_{i+1,x}\right].\nonumber
\end{equation}

We have shown that the unique nonzero optimal couplings are associated to the interactions $\hat\sigma_{i,x}\hat\sigma_{i+1,y}$ and $\hat\sigma_{i,y}\hat\sigma_{i+1,x}$ and are equal to $h$. In this way, taking into account that the ground state expectation values of Pauli operators with an odd number of $\hat\sigma_{i,x}$ or $\hat\sigma_{i,y}$ interactions vanish,  Eq.~\eqref{supp_2} becomes
\begin{equation}\label{supp_3}
	h=\frac{\langle \hat H_\text{p}\sum_n(\hat\sigma_{i,x}\hat\sigma_{i+1,y}+\hat\sigma_{i,y}\hat\sigma_{i+1,x})\rangle}{\langle\left( \sum_i(\hat\sigma_{i,x}\hat\sigma_{i+1,y}+\hat\sigma_{i,y}\hat\sigma_{i+1,x})\right)^2\rangle},
\end{equation}
where $\langle\cdot\rangle$ is the expectation value of an operator on the target state $\hat\rho(t)$.

\subsection{Pseudo-spin representation of the optimal coupling formula}

Equation~\eqref{supp_3} can be expressed using the pseudo-spin operators. Indeed,
\begin{align*}
	&\sum_i(\hat\sigma_{i,x}\hat\sigma_{i+1,y}+\hat\sigma_{i,y}\hat\sigma_{i+1,x})\\
	&\quad=2i\sum_i(\hat c_i^\dag \hat c_{i+1}^\dag+\hat c_i\hat c_{i+1})\\
	&\quad=2\sum_{k\in\mathcal{K}}(e^{ik}\hat c_k^\dag \hat c_{-k}^\dag-e^{-ik}\hat c_k\hat c_{-k})\\
	&\quad=4\sum_{k\in\mathcal{K}^+}\sin(k)i(\hat c_k^\dag \hat c_{-k}^\dag-\hat c_{-k}\hat c_k)\\
	&\quad=-4\sum_{k\in\mathcal{K}^+}\sin(k)\tilde\sigma_k^y,
\end{align*}
and, as shown in the main text,
\begin{equation}
	\hat H_\text{p}(t)=\frac{1}{2}\dot\lambda\sum_{k\in\mathcal{K}^+}\left(\partial_\lambda \theta_k\right)\tilde\sigma_k^y.
\end{equation}
Replacing these expressions in Eq.~\eqref{supp_3}, we obtain
\begin{equation*}
h=-\frac{\dot\lambda}{8}\frac{\sum_{k,k'\in\mathcal{K}^+}\langle \left(\partial_\lambda \theta_k\right)\tilde\sigma_k^y\sin(k')\tilde\sigma_{k'}^y\rangle}{\sum_{k,k'\in\mathcal{K}^+}\langle \sin(k)\tilde\sigma_k^y\sin(k')\tilde\sigma_{k'}^y\rangle}.
\end{equation*}

Finally, if we take into account the orthogonality relation $\langle \tilde\sigma_k^y\tilde\sigma_{k'}^y\rangle=\delta_{kk'}$, we obtain the final expression of the optimal coupling:
\begin{equation*}
	h=-\frac{\dot\lambda}{8}\frac{\sum_{k\in\mathcal{K}^+} \left(\partial_\lambda \theta_k\right)\sin(k)}{\sum_{k\in\mathcal{K}^+}\sin(k)^2}.
\end{equation*}

\section{Filtering out not allowed interactions}
\label{appendix_filter}

In this Appendix, we show that, given the time-dependent state $\hat\rho(t)$ and an exact parent Hamiltonian $\hat H_\text{p}(t):=\sum_k f_\alpha(t)\hat O_\alpha$ for this state, Eq.~(\ref{main_eq}) systematically defines a filter function $M_{a}^{\alpha}(t)$ such that the optimal parent Hamiltonian is
\be
\hat H_\text{opt}=\sum_ah_{\text{opt},a}(t)\hat L_{a},\nonumber
\ee
where $h_{\text{opt},a}(t)=\sum_\alpha M^\alpha_a(t)f_\alpha(t)$. This filter function defines new couplings for the allowed interaction in order to counterbalance, when possible, part of the effect of eliminating not allowed interactions.

We can derive the filter $M^\alpha_a(t)$ from Eq.~(\ref{main_eq}) by replacing the derivative of the state $\partial_t\hat\rho(t)$ with $-i[\sum_\alpha f_\alpha(t)\hat O_\alpha,\hat\rho(t)]$. In this way, we obtain
\be
h_{\text{opt},a}(t)=-\sum_{\alpha,a'}f_{\alpha}(V_{aa'}(t))^{-1}\Tr([\hat L_{a'},\hat\rho(t)][\hat O_{\alpha},\hat\rho(t)]),\nonumber
\ee
hence
\be\label{filter}
M^\alpha_{a}(t)=-\sum_{a'}(V_{aa'}(t))^{-1}\Tr([\hat L_{a'},\hat\rho(t)][\hat O_{\alpha},\hat\rho(t)]).
\ee
These last equations reflect the fact that the optimal parent Hamiltonian $\hat H_\text{opt}$ is the projection of an arbitrary exact parent Hamiltonian $\hat H_\text{p}$ on the space of the 
allowed interactions through the degenerate scalar product $(\cdot,\cdot)$ defined as $(\hat A,\hat B)=-\Tr([\hat A,\hat\rho][\hat B,\hat\rho])$. In the context of shortcuts to adiabaticity, this filter can also be used to systematically generate a local counterdiabatic potential starting from the exact one.

\section{Cost functional and counterdiabatic potential}
\label{appendix_polk}
Here, we show that the counterdiabatic potential defined in Ref.~\onlinecite{Polkovnikov} is equal to the potential that minimizes the functional in Eq.~\eqref{cost_functional} of the main text.

In Ref.~\onlinecite{Polkovnikov}, the counterdiabatic potential is defined as the operator $\hat A^*$ that minimizes
\begin{equation*}
	S_{\hat H_\text{a}}(\hat A^*)=\int_0^T \Tr\left[\bigl(\partial_t \hat H_\text{a}+i[\hat A^*,\hat H_\text{a}]\bigr)^2\right]dt,
\end{equation*}
where $\hat H_\text{a}(t)$ is the adiabatic Hamiltonian. The authors also show that the minima of the latter potential are also the minima of
\begin{equation*}
	S_{\hat H_\text{a}}''(\hat A^*)=\int_0^T \Tr\biggl[\Bigl(\partial_t \hat H_\text{a}+i[\hat A^*,\hat H_\text{a}]-\sum_i\partial_tE_i\hat \rho_i\Bigr)^2\biggr]dt,
\end{equation*}
where the $E_i$ and the $\hat \rho_i$ are the eigenvalues and the eigenvectors of $\hat H_\text{a}$ respectively. Exploiting the spectral decomposition of $\hat H_\text{a}$, this becomes
\begin{equation*}
S_{\hat H_\text{a}}''(\hat A^*)=\int_0^T \Tr\biggl[\Bigl(\sum_i E_i \bigl(\partial_t\hat \rho_i+i[\hat A^*,\hat \rho_i]\bigr)\Bigr)^2\biggr]dt.
\end{equation*}
Since this functional does not depend on the derivative of $\hat A^*$, its minimum is the minimum of the integrand function. Therefore, taking into account that the square root is a monotonic function, we can replace the integrand with its square root. We finally obtain
\begin{equation*}
S_{\hat H_\text{a}}'(\hat A^*)=\int_0^T \Bigl\lVert\sum_i E_i \bigl(\partial_t\hat \rho_i+i[\hat A^*,\hat \rho_i]\bigr)\Bigr\rVert dt,
\end{equation*}
where $\lVert\cdot\rVert$ is the Frobenius norm.


\end{document}